\documentclass[aps,prd,twocolumn,superscriptaddress,preprintnumbers,balancelastpage,longbibliography,nofootinbib]{revtex4-2}

\makeatletter
\def\@bibstyle{apsrev4-2-year}
\makeatother

\usepackage{appendix}
\usepackage[utf8]{inputenc}
\usepackage{amsmath,amssymb,mathtools,bm}
\usepackage{graphicx, color, hepunits}
\usepackage[dvipsnames]{xcolor}
\usepackage{float}
\usepackage{multirow}
 \usepackage{hyperref} 
\hypersetup{
    colorlinks=true,       
    linkcolor=blue,        
    citecolor=blue,        
    filecolor=magenta,     
    urlcolor=blue          
}
\usepackage[utf8]{inputenc}
\usepackage[english]{babel}

\usepackage{tensor}
\usepackage{booktabs} 
\newcommand{\aap}{Astron.\ Astrophys.}
\newcommand{\mnras}{Mon.\ Not.\ Roy.\ Astron.\ Soc.}

\newcommand{\es}[2] {\begin{equation} \label{#1} \begin{split} #2 \end{split} \end{equation}}

\begin{document}

\title{MeV dark matter decay forecasts for the COSI mission}

\author{Robin Anthony-Petersen}
\affiliation{Space Sciences Laboratory, 7 Gauss Way, University of California, Berkeley, CA 94720-7450, USA}

\author{Orion Ning}
\affiliation{Leinweber Institute for Theoretical Physics, University of California, Berkeley, CA 94720, USA}
\affiliation{Theoretical Physics Group, Lawrence Berkeley National Laboratory, Berkeley, CA 94720, USA}

\author{Yu Watanabe}
\affiliation{Department of Physics and Astronomy, University of California, Los Angeles, CA 90095, USA}

\author{Benjamin R. Safdi}
\affiliation{Leinweber Institute for Theoretical Physics, University of California, Berkeley, CA 94720, USA}
\affiliation{Theoretical Physics Group, Lawrence Berkeley National Laboratory, Berkeley, CA 94720, USA}

\author{Maíra Dutra}
\affiliation{Southeastern Universities Research Association (SURA), 1201 New York Avenue NW, Suite 430, Washington, DC 20005, USA}
\affiliation{NASA Goddard Space Flight Center, 8800 Greenbelt Road, Greenbelt, MD 20771, USA}

\author{Francesca Calore}
\affiliation{Laboratoire d'Annecy de Physique Th\'eorique (LAPTh), CNRS, USMB, 74940 Annecy, France}

\author{Savitri Gallego}
\affiliation{Institut für Physik \& Exzellenzcluster PRISMA\textsuperscript{++}, Johannes Gutenberg-Universität Mainz, 55099 Mainz, Germany}

\author{William Guan}
\affiliation{Harvard University, Cambridge, Massachusetts 02138, USA}

\author{Christopher M. Karwin}
\affiliation{Department of Physics and Astronomy, Clemson University, Clemson, SC 29634, USA}

\author{Alexander Kusenko}
\affiliation{Department of Physics and Astronomy, University of California, Los Angeles, CA 90095, USA}
\affiliation{Kavli Institute for the Physics and Mathematics of the Universe (IPMU, WPI), The University of Tokyo, 5-1-5 Kashiwanoha, Kashiwa, Chiba 277-8583, Japan}

\author{Israel Martinez-Castellanos}
\affiliation{Department of Astronomy, University of Maryland, College Park, MD 20742, USA}
\affiliation{NASA Goddard Space Flight Center, 8800 Greenbelt Road, Greenbelt, MD 20771, USA}

\author{Shigeki Matsumoto}
\affiliation{Kavli Institute for the Physics and Mathematics of the Universe (IPMU, WPI), The University of Tokyo, 5-1-5 Kashiwanoha, Kashiwa, Chiba 277-8583, Japan}

\author{Michela Negro}
\affiliation{Department of Physics and Astronomy, Louisiana State University, Baton Rouge, LA 70803, USA}

\author{Uwe Oberlack}
\affiliation{Institut für Physik \& Exzellenzcluster PRISMA\textsuperscript{++}, Johannes Gutenberg-Universität Mainz, 55099 Mainz, Germany}

\author{Field R. Rogers}
\affiliation{Space Sciences Laboratory, 7 Gauss Way, University of California, Berkeley, CA 94720-7450, USA}

\author{Thomas Siegert}
\affiliation{Julius-Maximilians-Universität Würzburg, Fakultät für Physik und Astronomie, Institut für Theoretische Physik und Astrophysik,Lehrstuhl für Astronomie, Emil-Fischer-Str. 31, D-97074 Würzburg, Germany}

\author{Tadayuki Takahashi}
\affiliation{Kavli Institute for the Physics and Mathematics of the Universe (IPMU, WPI), The University of Tokyo, 5-1-5 Kashiwanoha, Kashiwa, Chiba 277-8583, Japan}

\author{Haoyu Xie}
\affiliation{Institut für Physik \& Exzellenzcluster PRISMA\textsuperscript{++}, Johannes Gutenberg-Universität Mainz, 55099 Mainz, Germany}

\author{Hiroki Yoneda}
\affiliation{Department of Physics, Kyoto University, Kitashirakawa Oiwake-cho, Sakyo-ku, Kyoto 606-8502, Japan}
\affiliation{RIKEN Nishina Center, 2-1 Hirosawa, Wako, Saitama 351-0198, Japan}
\affiliation{Kavli Institute for the Physics and Mathematics of the Universe (IPMU, WPI), The University of Tokyo, 5-1-5 Kashiwanoha, Kashiwa, Chiba 277-8583, Japan}

\author{Andreas Zoglauer} 
\affiliation{Space Sciences Laboratory, 7 Gauss Way, University of California, Berkeley, CA 94720-7450, USA}

\author{Steven E. Boggs} 
\affiliation{Department of Astronomy \& Astrophysics, University of California, San Diego, 9500 Gilman Drive, La Jolla, CA 92093-0424, USA}

\author{Dieter H. Hartmann} 
\affiliation{Department of Physics and Astronomy, Clemson University, Clemson, SC 29634, USA}

\author{Carolyn A. Kierans} 
\affiliation{NASA Goddard Space Flight Center, 8800 Greenbelt Road, Greenbelt, MD 20771, USA}

\author{John A. Tomsick} 
\affiliation{Space Sciences Laboratory, 7 Gauss Way, University of California, Berkeley, CA 94720-7450, USA}

\date{\today}

\begin{abstract}
The decay of dark matter (DM) is a generic prediction of many particle physics models, producing distinctive monochromatic signatures in environments such as the Galactic halo. Searches for decaying DM in the elusive `MeV gap' have remained conspicuously under-explored, despite well-motivated models including axion-like particles, sterile neutrinos, and hidden-sector scenarios predicting signatures in this energy range. The upcoming Compton Spectrometer and Imager (COSI) telescope will help bridge this gap. Designed to probe this energy band with high spectral resolution, low background, and wide field of view, COSI is well-suited for DM line searches at the MeV scale. In this work, we make projections for COSI's 2-year sensitivity for DM decay in the Milky Way halo using dedicated all-sky simulations incorporating COSI's predicted instrument response. We conclude that COSI will improve the sensitivity to the DM lifetime by more than an order of magnitude over the DM mass range 0.4 to 10 MeV.
\end{abstract}

\maketitle

\section{Introduction} 
\label{sec:intro}

Dark matter (DM) decay to Standard Model final states is a generic expectation of particle DM scenarios because quantum gravity is expected to violate global symmetries~\cite{Kallosh:1995hi,Banks:2010zn,Harlow:2018tng} and the Standard Model contains massless and ultralight particles ({\it e.g.}, the photon and neutrinos). 
Thus, unless the DM candidate is charged under a discrete or continuous gauge symmetry, the effective field theory expectation is that DM may decay through higher-dimensional operators suppressed by physics at or below the Planck scale. While DM must be stable on cosmological timescales~\cite{Kolb:1990vq,Clowe:2006eq,Garrett:2010hd,Dodelson:2011qv}, modern telescopes probe lifetimes that are some $10^{10}$--$10^{12}$ times longer and that encroach on the expectations of explicit particle DM scenarios.  Particle DM models that include potentially-detectable DM decay in the MeV mass range include axion-like particle DM~\cite{Arvanitaki:2009fg,Cadamuro:2011fd,Foster:2022ajl}, sterile neutrino DM~\cite{Dodelson:1993je,Kusenko:2009up,Fujisawa:2025yqi}, hidden-sector glueball DM~\cite{Juknevich:2009gg,Cohen:2016uyg,Acharya:2017szw}, among other models~\cite{Essig:2013goa,Kusenko:2012ch, Feldstein:2013kka}. Moreover, such models typically have sizable or dominant branching ratios to monochromatic final states ($\gamma + X$, with $X$ a photon or neutrino).

\begin{figure}[t]
   \centering
   \includegraphics[width=1.0\linewidth]{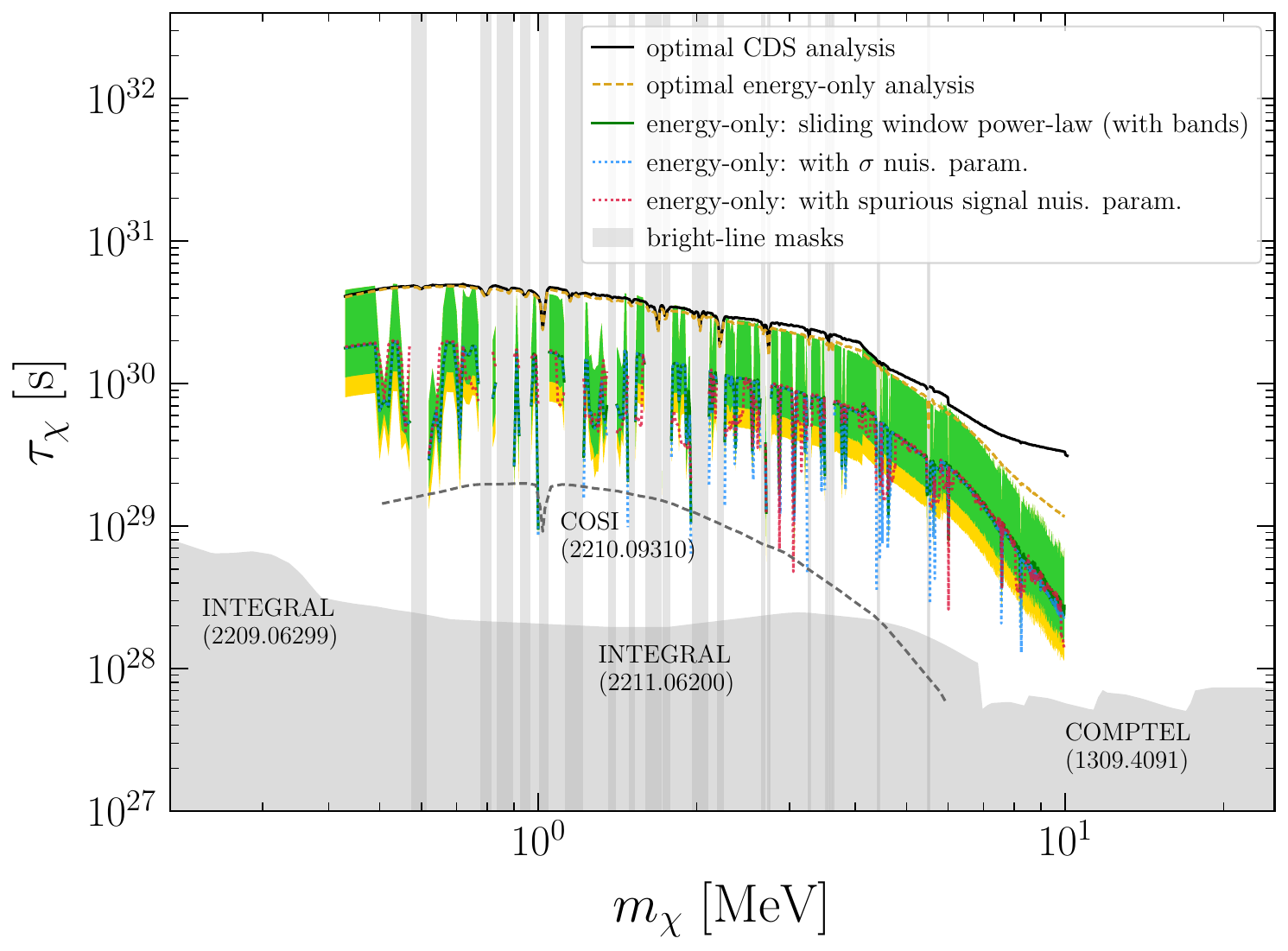}
   \caption{Projected 95\% confidence level lower limits on the DM decay lifetime, under the null hypothesis, as a function of the DM mass assuming two years of COSI data. The results assume a monochromatic $\gamma$-ray line signal, an NFW DM profile in the Milky Way, and are derived using binned likelihood analyses of simulated COSI observations.  We show the projected sensitivity for the optimal Compton Data Space (CDS) analysis and a simplified analysis that only uses spectral and not spatial data. We then illustrate how the sensitivity is degraded by mock analyses that are more realistic, incorporating an imperfect power-law background model over a sliding energy window (we also show the expected $1/2\sigma$ containment windows for the power-constrained limit under the null hypothesis for this analysis) and then further incorporating a systematic nuisance parameter for mismodeling that inflates the statistical error bars and a systematic nuisance parameter, the spurious signal, that works to achieve the correct distribution of test statistics (TSs). For these realistic analyses, projections are computed using the Asimov data set determined by fits to the 3-month Fourth Data Challenge (DC4) mock data  (see text for details). For comparison, existing constraints from INTEGRAL/SPI~\cite{Calore:2022pks,Fischer:2022pse} and COMPTEL~\cite{Essig:2013goa} are shown in gray. The limits of Ref.~\cite{Fischer:2022pse} are rescaled to account for a two-photon final state, and all existing limits are otherwise shown as published with their own DM halo parameters, given that these enter non-trivially through analysis-specific spatial templates and sky regions. Also shown are initial COSI projections~\cite{Caputo:2022dkz} utilizing just the estimated point source line sensitivity and without the full instrument response framework. COSI is expected to significantly improve sensitivity across the MeV mass range, probing DM lifetimes up to $\mathcal{O}(10^{30})~\mathrm{s}$, even accounting for realistic analysis variants.
   }
   \label{fig:cosi_limits}
\end{figure}

Searches for photon lines from DM decay have been performed across the electromagnetic spectrum. At keV energies, data from XMM-Newton, Chandra, NuSTAR, Suzaku, and eROSITA instruments have yielded strong limits on the DM lifetime, excluding lifetimes $\tau_\chi \lesssim  10^{28} - 10^{29}$~s for DM masses roughly between 1 and 50 keV (see, \textit{e.g.},~\cite{Loewenstein:2008yi,Loewenstein:2009cm,Loewenstein:2012px,Roach:2019ctw,Foster:2021ngm,Dessert:2023fen,Roach:2022lgo, TerolCalvo:2026nlr}). At GeV energies and above, the Fermi Large Area Telescope (LAT) and ground-based Cherenkov arrays have constrained DM lifetimes up to $10^{30}$~s~\cite{Zhang:2009ut,Cohen:2016uyg,Foster:2022nva}. Between these two regimes lies the MeV gap, spanning roughly ${\sim}0.1$--$100$ MeV, where the only existing line constraints come from the decades-old instruments INTEGRAL/SPI~\cite{Calore:2022pks,Fischer:2022pse} and COMPTEL~\cite{Essig:2013goa}. In this work we will show that the upcoming Compton Spectrometer and Imager (COSI) space-based telescope mission will enhance the sensitivity to decaying DM into two photons by over an order of magnitude compared to previous constraints from INTEGRAL and COMPTEL in the mass range between roughly 0.4 to 10 MeV (see Fig.~\ref{fig:cosi_limits}).

A NASA Small Explorer (SMEX) mission scheduled to launch in August 2027, with a nominal mission duration of two years, COSI will survey the entire sky between ${\sim}0.2$ and $5$ MeV, measuring both the energies and arrival directions of incident $\gamma$-rays~\cite{2024icrc.confE.745T}. Compared with previous MeV $\gamma$-ray missions, COSI is set apart by the combination of its large instantaneous field of view (FOV) ($>$25\% of the sky), excellent energy resolution ($\mathcal{O}(1\%)$ near 1 MeV), and low instrumental backgrounds. These features make COSI particularly well suited to search the Milky Way halo for diffuse line features.

Note that COSI's expected improvement over INTEGRAL can be roughly estimated using each telescope's FOV, observation time, and background rates. In two years, COSI's wide FOV will enable the detection of ${\sim}1.5-4\times$ the number of photons INTEGRAL detected in 16 years, approximately the observation time used to derive the limit in Ref.~\cite{Fischer:2022pse} and shown in Fig.~\ref{fig:cosi_limits}. In addition, comparing COSI's pre-flight projected backgrounds to the measured INTEGRAL backgrounds from Fig. 2 in Ref.~\cite{2018A&A...611A..12D}, COSI has ${\sim}25 \times$ lower backgrounds than INTEGRAL. While the exact background ratio must be updated once COSI is on orbit, COSI is expected to have overall lower backgrounds, a result of using Compton event reconstruction, the anti-coincidence shield vetoes, as well as design choices which reduce active material around the detector array. Combined, this gives an expected back-of-the-envelope improvement of $(1.5-4) \times (\sqrt{25}) = {\sim}8-20$ for COSI relative to INTEGRAL. However, this scaling assumes that both instruments are statistics limited, which is not the case for INTEGRAL, whose line limits are systematics dominated~\cite{2018A&A...611A..12D,Fischer:2022pse}. In this work we account for the most up-to-date COSI simulations and realistic analysis frameworks that address systematics mismodeling to make an accurate projection for COSI's sensitivity.

Our projections are obtained by propagating simulated DM decay signals through a detailed model of the COSI instrument response, together with dedicated simulations of the dominant astrophysical and instrumental backgrounds expected during the mission. This is in contrast to previous work, see {\it e.g.} Ref.~\cite{Caputo:2022dkz}, which estimated COSI's DM sensitivity by making projections based on the published line sensitivity for point sources from Ref.~\cite{2022icrc.confE.652T}, without the full forward modeling framework of the COSI instrument response. Accounting for the full instrument response, and extending our analysis to include the entire sky, we show that the sensitivity to the DM lifetime is over an order of magnitude better than anticipated in Ref.~\cite{Caputo:2022dkz} for a range of DM masses.

The remainder of this work is organized as follows. Sec.~\ref{sec:cosi} describes the COSI instrument. Sec.~\ref{sec:modeling_analysis} introduces our signal and background models and our proposed analysis frameworks. We present our projected sensitivities in Sec.~\ref{sec:results} and discuss the implications for MeV-scale decaying DM in Sec.~\ref{sec:discussion}.


\section{The COSI Instrument}
\label{sec:cosi}

COSI detects MeV $\gamma$-rays via Compton scattering. For this, it employs an array of 16 high-purity cross-strip germanium detectors (GeDs) surrounded on the sides and bottom by an anti-coincidence shield. As a $\gamma$-ray in the COSI band traverses the detector volume, it preferentially Compton scatters, progressively losing energy until being photoabsorbed. At each interaction site, the energy deposition and position are measured, and the incident $\gamma$-ray is parameterized by each of its interactions and their respective energy and positional measurements. 

COSI's spectroscopic and imaging capabilities are inherently related to Compton reconstruction. With the information of a single $\gamma$-ray distributed across all its interaction sites, the total energy deposition is obtained by summing the measured energies at each interaction. This is how COSI performs spectroscopy. The energy resolution requirement is a full width at half maximum (FWHM) of $<$6.0~keV at $0.511$~MeV and $<$9.0~keV at $1.157$~MeV. For imaging, knowing the origin of the incoming $\gamma$-ray is required. COSI infers this information by using the kinematics and geometry of the first two interaction sites to measure the Compton scattering angle, $\phi$, thereby constraining the polar scattering angle. However, because COSI cannot precisely measure the direction of the Compton scatter recoil electron, this leaves the azimuthal degree of freedom unconstrained. As a result, the arrival direction of a $\gamma$-ray cannot be constrained to a point, but instead it becomes defined by a cone with opening angle $\phi$ and center (after a rotation to transform the local axis into Galactic coordinates) at ($\ell$, $b$). The origin of the $\gamma$-ray then lies along the projection of this cone onto the sky~\citep{2018PhDT.......171K,Kierans:2022eid}. 

The Compton cone structure represents a logical choice for a Compton telescope's data space, which we refer to as the Compton Data Space (CDS) (first formalized by COMPTEL in Ref.~\cite{1992NASCP3137...95D}). In the CDS, each event becomes a single point defined by the coordinates of the Compton cone ($\phi$, $\ell$, $b$) and the total deposited energy, $E$. To reconstruct an all-sky spectrum we simply sum over the angular components. Keeping the angular coordinates provides imaging information, allowing for the spatial structures of different sources to be probed, and yields improved background rejection. A more detailed formulation of the CDS and its application to this analysis are discussed in Sec.~\ref{sec:cds}.

\section{Modeling and Analysis}
\label{sec:modeling_analysis}

In this section, we cover how our projected 95\% confidence-level lower limits on $\tau_{\chi}$ are calculated. We describe the modeling of the DM signal (Sec.~\ref{sec:dm_model}) and the simulation pipeline used to forward model our DM signal and the background data (Sec.~\ref{sec:sig_bkg}). With these simulations, we forecast the sensitivity using two approaches. First, we use the reconstructed energy spectrum, summed over the FOV, to conduct an energy-only line search (Sec.~\ref{sec:energy_only}). We present this as our main result since the background can be parameterized and profiled, enabling the limit to account for background uncertainty. Second, we consider the full CDS, which adds spatial information (Sec.~\ref{sec:cds}). For the CDS analysis, the background is treated as fixed because the single available background realization is too sparse in four dimensions to constrain free parameters meaningfully. This means the systematic mismodeling uncertainties are not modeled and this limit represents a best-case scenario, bounding the gain in sensitivity available from the spatial information. 

\subsection{Dark Matter Signal Model}
\label{sec:dm_model}

The differential photon flux per unit solid angle $\frac{d^2 \Phi}{dE\, d\Omega}$ defines the expected photon rate and spatial morphology of the signal from DM decay in the Galactic halo~\cite{Safdi:2022xkm}: 
\begin{equation}
\label{eq:signal_flux}
    \frac{d^2 \Phi}{dE\, d\Omega} (E, \ell, b) = \frac{N_\gamma}{4 \pi\, m_\chi\, \tau_\chi}\, \delta(E - m_\chi/2)\, \mathcal{D}(\ell, b) \,,
\end{equation}
where $N_\gamma = 2$ is the number of photons produced per decay,
$\tau_\chi$ is the partial lifetime to two photons, and $\mathcal{D}(\ell, b)$ is the astrophysical $D$-factor along the line-of-sight at Galactic coordinates $(\ell, b)$,
\begin{equation}
\label{eq:d-factor}
    \mathcal{D}(\ell, b) = \int_{\rm l.o.s.} \! ds\, \rho_{\rm DM}(r) \,,
\end{equation}
with $\rho_{\rm DM}(r)$ the DM density at Galactocentric distance $r$ and $s$ the line-of-sight distance. For definiteness we assume DM decays of the form $\chi \to \gamma + \gamma$, though our results are straightforwardly extended to $\chi \to \gamma + X$, with $X$ a neutral final-state particle, by taking $N_\gamma = 1$ and replacing $m_\chi/2 \to E_\gamma = (m_\chi/2)(1 - m_X^2/m_\chi^2)$ in Eq.~\eqref{eq:signal_flux}. For $\chi \to \gamma + \gamma$ the photon spectrum is monochromatic with $E_{\gamma} = m_{\chi}/2$ and an intrinsic width set by the DM velocity dispersion in the Milky Way, $\delta E / E \sim v/c \sim 10^{-3}$. Compared to COSI's energy resolution of $\mathcal{O}(1\%)$ near $1$~MeV, the width of the DM line is taken to be negligible.

We model $\rho_{\rm DM}$ with a Navarro--Frenk--White (NFW) profile~\cite{1996ApJ...462..563N,1997ApJ...490..493N},
\begin{equation}
\label{eq:NFW}
    \rho_{\rm DM}(r) = \frac{\rho_s}{(r/r_s)\,(1+r/r_s)^2} \,.
\end{equation}
We follow the prescription of Ref.~\cite{Foster:2021ngm} and fix $\rho_s = 0.26~{\rm GeV/cm^3}$ and $r_s = 19.1~{\rm kpc}$.  These are the most conservative values within the $68\%$ uncertainty contour reported in Ref.~\cite{2020MNRAS.494.4291C}, which fits the Milky Way mass profile using Gaia DR2 rotation curve data and satellite kinematics. This choice yields a corresponding local DM density of $\rho_\odot \simeq 0.29~{\rm GeV/cm^3}$ at the solar radius $r_\odot = 8.23~{\rm kpc}$.

\subsection{Signal and Background Simulations}
\label{sec:sig_bkg}

We forward-model the signal in Eq.~\eqref{eq:signal_flux} through the COSI instrument response using the Medium-Energy Gamma-ray Astronomy library (MEGAlib)~\cite{2006NewAR..50..629Z}, which provides a GEANT4-based model of the COSI instrument. Monochromatic photons with $E = m_{\chi}/2$, over an array of possible $m_\chi$, are sampled from the NFW $\mathcal{D}$-factor map in Eq.~\eqref{eq:d-factor} and propagated through the instrument model.  

\begin{figure}[h]
   \centering
   \includegraphics[width=1.0\linewidth]{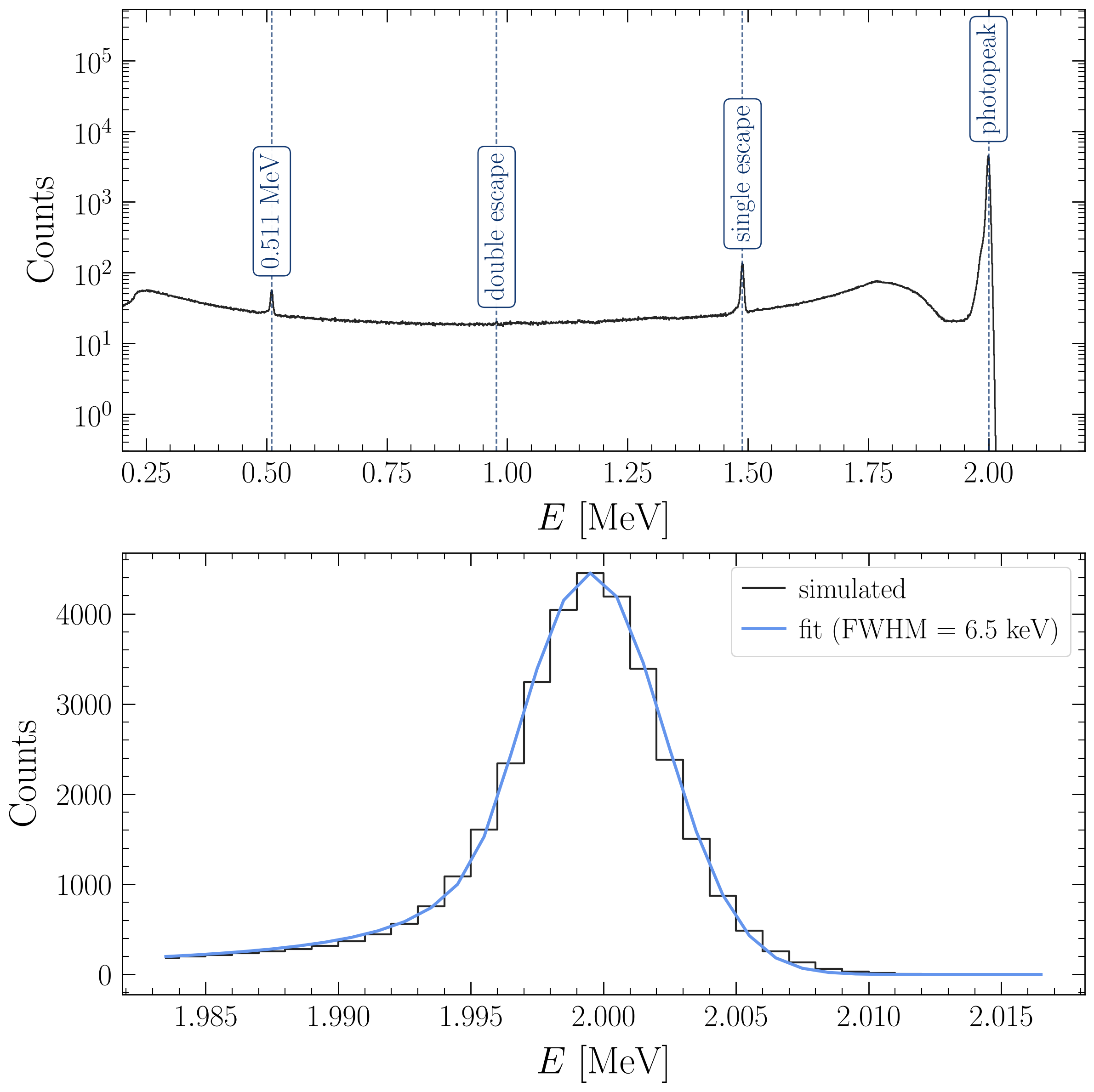} 
   \caption{Forward-modeled energy spectrum for a simulated $\chi \to \gamma + \gamma$ signal with $\tau_{\chi} = 10^{28}$ s at $m_\chi=4$~MeV ($E_\gamma=2$~MeV), averaged over $100$ MC realizations, with 3 months of exposure. \emph{Top:} the full spectrum, with the signal model summed over the angular CDS coordinates, showing the full-energy photopeak at $2.000$~MeV, the single- and double-escape peaks at $1.489$ and $0.978$~MeV (loss of one or both $0.511$~MeV annihilation photons after pair production), the $0.511$~MeV annihilation line, and the Compton continuum below the photopeak. \emph{Bottom:} The photopeak fit to determine the sliding analysis energy window (FWHM $6.5$~keV in this case).}
   \label{fig:spectrum}
\end{figure}

Our data space is discretized into 1~keV energy bins and, in the CDS, $15^\circ$ bins in each of the Galactic coordinates ($\ell$, $b$) of the Compton cone's axis and the Compton scattering angle $\phi$. We generate signal templates at 971 mass points spanning $m_\chi \in [0.4, 10.1]~{\rm MeV}$, with $10$~keV spacing ($5$~keV in the photon energy $E_\gamma = m_\chi/2$), finer than the COSI energy resolution across the band. At each mass point, we generate 100 independent Monte Carlo (MC) realizations of the forward-modeled signal, assuming $\tau_\chi = 10^{28}$~s, in the data space and average over them to suppress MC noise, yielding the expected number of counts per bin. An example spectrum is shown in Fig.~\ref{fig:spectrum}.

Projections of the expected instrumental and astrophysical backgrounds are provided by the COSI collaboration and updated yearly as part of their public Data Challenges~\cite{cosi_dc_2026}. In our analysis, we use the mock datasets prepared for the fourth COSI Data Challenge (DC4), which simulates three months of observations. They contain the total background set of emissions described in Ref.~\cite{2026ApJ...997..284G} which are common across multiple science analyses, as well as 64 astrophysical sources. The total background components are Galactic diffuse continuum components (notably inverse Compton radiation and Bremsstrahlung), extragalactic $\gamma$-ray background, albedo photon emission, South Atlantic Anomaly (SAA) protons, and prompt and delayed instrumental backgrounds arising from proton, electron, alpha-particle, positron, and albedo neutron activation. The astrophysical sources include positron, nucleosynthesis ({\it e.g.}\ $^{26}$Al, and $^{60}$Fe), Galactic, extragalactic, and transient ({\it i.e.}\ $\gamma$-ray bursts) sources. A complete list of the simulated sources, along with details on their input parameters, can be found in Ref.~\cite{Karwin_inprep}, or on the COSI Data Challenge GitHub (see Ref.~\cite{cosi_data_challenges} for more information about the sources and Ref.~\cite{cosi_dc4_dataproducts} for the mock dataset simulation files).

The signal and background events are generated with the same GEANT4-based COSI instrument model and processed with the COSI Collaboration's standard event-reconstruction and selection cuts (see Refs.~\cite{zoglauer2021cosicalibrationsobservationsallsky} for more details on the COSI analysis pipeline). This ensures that the event distributions from both are represented consistently at the detector-level in the subsequent analyses. 

Both the signal and background simulations correspond to a three-month exposure. The resulting projected limits are scaled to the two-year baseline COSI mission time by observing that in the background-limited regime the lifetime limit scales as $\sqrt{\mathrm{exposure}}$, such that the rescaling factor is $\sqrt{24/3}\approx2.8$, assuming the background level stays constant throughout the mission time.\footnote{The assumption that the background level stays constant is reasonable for the continuum. This does not hold for the instrumental activation lines, which increase over the mission's lifetime. However, because the bright lines are masked and the dim lines are modeled with free normalizations (Sec.~\ref{sec:energy_only}), this likely does not significantly alter our analysis.} 

\subsection{Energy-Only Analysis}
\label{sec:energy_only}

In this section, we consider a spectral analysis based only on reconstructed photon energies, summed over the angular CDS coordinates.\footnote{Note that there is no selection for region of interest in this approach: we consider all photons observed across the full sky. Appendix~\ref{app:ROI} explores the effects on the sensitivity from selecting events from certain sky regions.}
At a fixed DM mass $m_\chi$, we model the expected counts in the data-space energy bin $i$ as the sum of a signal and background contribution,
\begin{equation}
    \mu_i(m_{\chi}, \tau_\chi, \boldsymbol{\theta}) = s_i(m_{\chi}, \tau_\chi) + b_i(\boldsymbol{\theta}) \,,
\end{equation}
where $s_i(m_\chi, \tau_\chi)$ is obtained by forward-modeling the signal in Eq.~\eqref{eq:signal_flux} through the COSI instrument response, with $s_i \propto \tau_\chi^{-1}$, and $b_i(\boldsymbol{\theta})$ is the background prediction. The vector $\boldsymbol{\theta}$ denotes the background-model nuisance parameters, which are described in more detail below. Given that the expected number of counts in each data-space energy bin is large (see, {\it e.g.}, Fig.~\ref{fig:spectrum}), we are justified in making a Gaussian approximation to the Poisson likelihood:
\begin{equation}
\begin{split}
    -2 \ln \mathcal{L}\big({\bf d}; {\mathcal M}, \{m_\chi, \tau_\chi, \boldsymbol{\theta}\}\big) ={}\\
    \sum_i \bigg[ \frac{\big(n_i - \mu_i\big)^2}{\sigma_i^2} + \ln\!\big(2\pi \sigma_i^2\big) \bigg] \,.
\end{split}
\label{eq:likelihood}
\end{equation}
where ${\bf d} = \{n_i\}_i$ is the set of observed counts over energy bins, $\sigma_i$ is the uncertainty associated with the observed counts ($\sigma_i = \sqrt{n_i}$ for a Poisson likelihood), ${\mathcal M}$ denotes the DM model with signal ($m_\chi, \tau_\chi$) and nuisance $({\bm \theta})$ parameters, and the sum runs over all bins within the analysis energy window, which is discussed more below.

Given a background model and analysis window, we project the expected sensitivity to the DM lifetime at fixed DM mass $m_\chi$ by computing the profile likelihood with the Asimov data set~\cite{Cowan:2010js} (see Ref.~\cite{Safdi:2022xkm} for a review). In particular, we compute the 95\% one-sided, expected, power-constrained lower limits to $\tau_\chi$ under the null hypothesis~\cite{Cowan:2011an}.  We profile over the full set of background-model nuisance parameters ${\bm \theta}$.

Below, we perform two different sets of analyses. First, we present results for the sensitivity projection of COSI, assuming no mismodeling. Second, we use the simulated background data and a realistic analysis framework to estimate the effect of background mismodeling on actual data.

\subsubsection*{Sensitivity projections}

We perform multiple sensitivity projections with increasing complexity of the background modeling to assess both the DM sensitivity and possible confounding sources of systematic uncertainty. First, in Fig.~\ref{fig:cosi_limits} we show the optimal sensitivity to a DM signal in the energy-only analysis (`optimal energy-only analysis'), which is calculated by having no selection on the analysis energy range for a given $m_\chi$ and assuming that the background model $b_i$ is a perfect description of the true background, with no nuisance parameters. A realistic analysis will have nuisance parameters and thus a sensitivity to the DM lifetime that is strictly less than this sensitivity.  

As a slightly more realistic analysis framework, we use a narrow, sliding analysis energy window and a parametric background model.  More specifically, we take the sliding energy window to be $\pm 2 \sigma_E$ centered on $m_\chi / 2$ for a given $m_\chi$, with $\sigma_E \equiv {\rm FWHM} / (2 \sqrt{2 \ln 2})$ and FWHM the full width at half maximum of the COSI energy response at that energy.\footnote{For this analysis, we determine the FWHM by fitting the DM simulated lines, as shown for one example line in Fig.~\ref{fig:spectrum}. These points are then interpolated across COSI's $0.2-5$~MeV band pass to determine $\sigma_E(E)$ in keV, which increases with energy ({\it e.g.} $\sigma_E(0.4$~MeV) = 2.1~keV, $\sigma_E(2$~MeV) = 2.8~keV).}  Within this sliding window we model the continuum data by a power-law with spectral index and overall normalization treated as independent nuisance parameters. Additionally, we model and mask astrophysical and instrumental lines, as described more below.

We identify background lines using our sliding window analysis over the simulated background, fitting a local power-law continuum plus a line profile to determine a candidate line's significance. This quantifies the brightness of a line, and yields a data-driven catalog of line energies and significances, for which we select bright lines with significance $\sigma_l \geq 30$ to be masked (for details on this procedure, see App.~\ref{app:bkg_lines}).  Energy bins within $\pm 5 \sigma_E$ of the bright lines are masked. The dim lines, which are detected with $\sigma_l < 30$, are not masked but rather are modeled in the analysis on top of the power-law continuum background model.\footnote{Masked and modeled line energies are illustrated in App.~\ref{app:bkg_lines} Fig.~\ref{fig:background_lines}, as determined from the 3 months of simulated background data.} In particular, each dim line is given a nuisance parameter for its normalization.  Note that we include dim lines that are outside of the energy range of interest but within $1 \sigma_E$ of the analysis-window edge.  Thus, if $N$ background lines are allowed by the line selection criterion, the background model has $N + 2$ model parameters, which are all profiled over when constructing the profile likelihood for $\tau_\chi$.

It is important to emphasize that the above procedure introduces a number of ad-hoc choices, such as the choice of which lines to mask or model and the parametric power-law continuum. The question that we ask is if this set of choices leads to an unbiased sensitivity to a DM signal across the mass range, with the distribution of discovery test statistics (TSs), for example, consistent with the expectation under the null hypothesis, as we discuss more below. This approach to continuum and line modeling is standard in DM-induced line searches; our approach, in particular, is modeled after that in Ref.~\cite{Foster:2021ngm}.

To assess the degradation in sensitivity of the sliding-window analysis relative to the optimal analysis, we first assume no mismodeling through the following procedure. First, we fit the background model to the three months of simulated background data, using the full sliding window analysis with the inclusion of the bright line masks and dim line model components. We then assume that the best-fit background model is the Asimov data set for a new analysis, at that same $m_\chi$ point, and compute the expected 95\% one-sided upper limit on $\tau_\chi^{-1}$. By construction, this analysis thus has no mismodeling. The projected 95\% lower limit on $\tau_\chi$ from this analysis is shown in Fig.~\ref{fig:cosi_limits}, scaled to the two-year sensitivity (`energy only: sliding window power-law'). The sensitivity is degraded relative to the optimal analysis by around a factor of two or more across the full unmasked mass range of interest, with the sensitivity degradation becoming more pronounced at higher masses.

\subsubsection*{Assessing effects of background mismodeling}

\begin{figure*}[htbp] 
   \centering
   \includegraphics[width=0.48\textwidth]{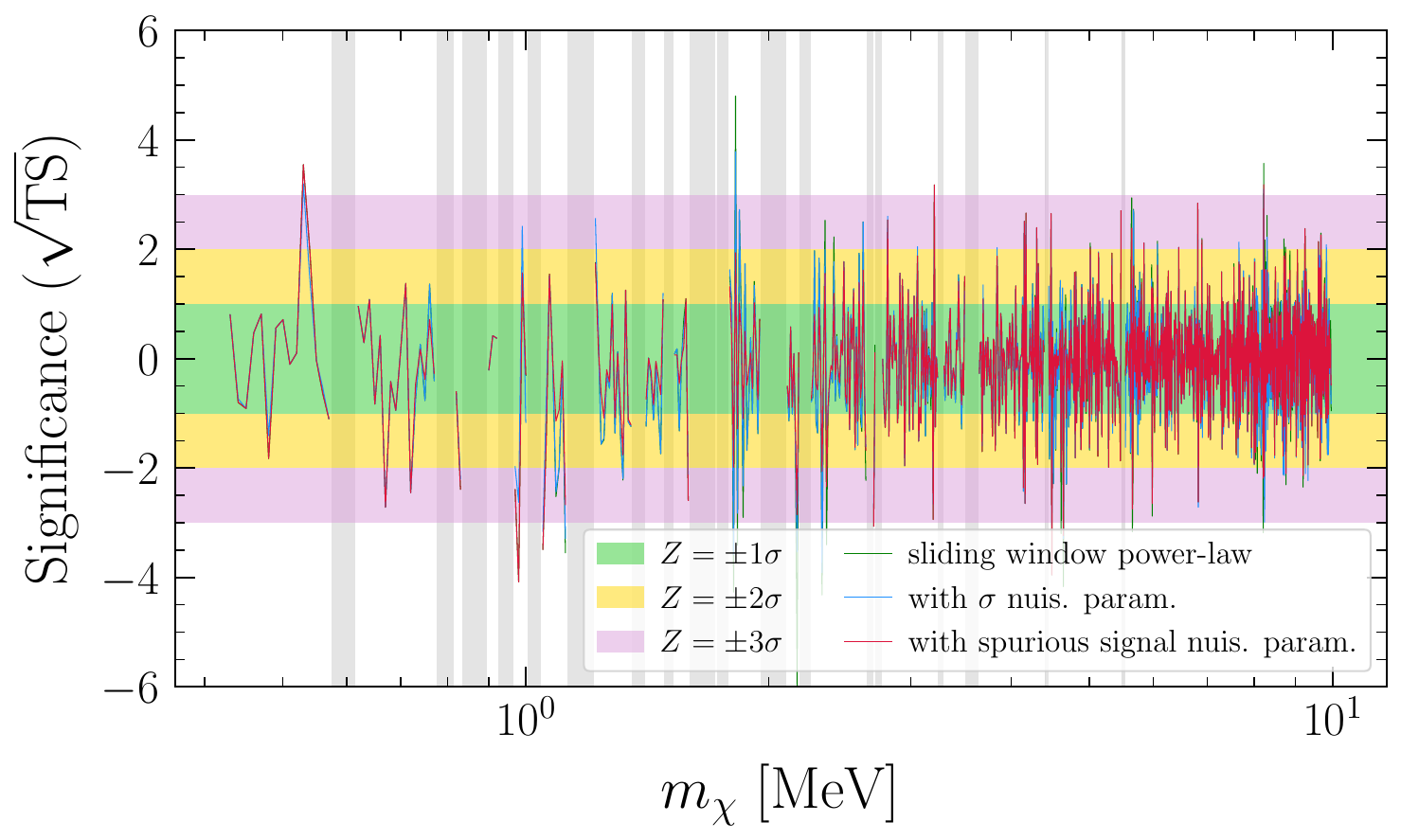}
   \includegraphics[width=0.48\textwidth]{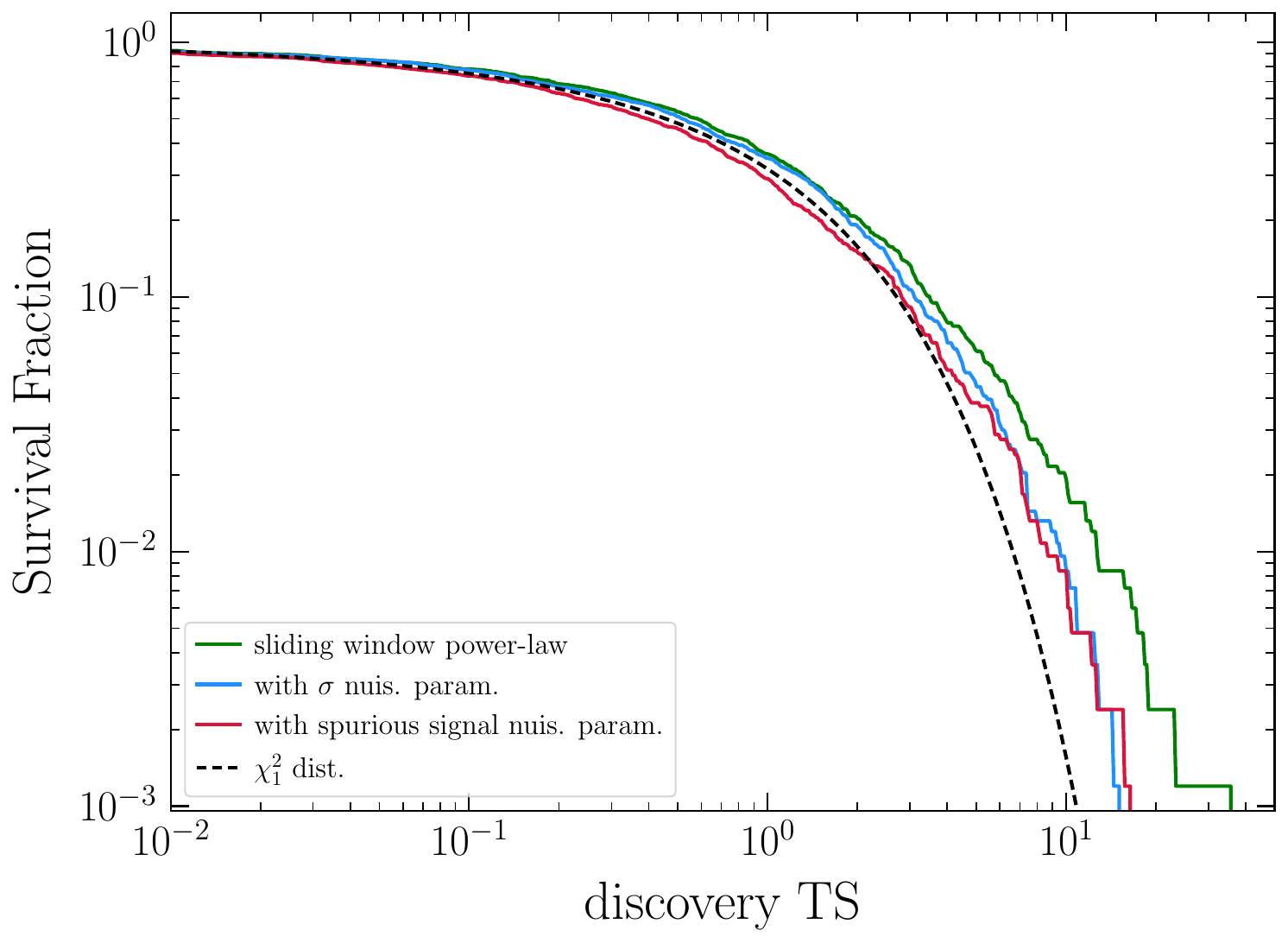}
   \caption{\emph{Left:} The sign-weighted significance ($Z$) in favor of the DM model as a function of DM mass, across our various spectral analysis frameworks, with $|Z| = 1, 2, 3$ markers. \emph{Right:} The discovery TS survival fraction over all unmasked masses for the same models, compared to the ideal $\chi^2_1$ distribution.}
   \label{fig:ts_variants}
\end{figure*}

The sliding-energy-window power-law-based analysis is meant to mitigate the effects of background mismodeling, acknowledging the reality that we will not have a perfect background model description and will need to incorporate parametric or non-parametric components at some level in a realistic analysis. To assess the effectiveness of this approach, we apply this analysis procedure to the simulated 3 months of null-hypothesis data. In Fig.~\ref{fig:ts_variants} (left panel) we show the discovery TS in favor of the DM model times the sign of the best-fit signal strength (denoted $Z$), as a function of DM mass. (Note that we consider negative signal strengths, even though they are unphysical, for statistical consistency~\cite{Safdi:2022xkm}.) The locations of masked energy bins are shown in gray, and we do not consider DM masses that fall within masked bins.  The 1$\sigma$, 2$\sigma$, and 3$\sigma$ expectations for the discovery TS under the null hypothesis are given by the green, gold, and purple bands, respectively, and various high-TS excesses and deficits are observed. In the right panel of Fig.~\ref{fig:ts_variants} we show the survival fraction of discovery TSs, compared to the expectation under the null-hypothesis given by the survival fraction of the chi-square distribution with one degree of freedom. The analysis overshoots that null-hypothesis expectation at large TS values, indicating evidence for mismodeling.

To further mitigate mismodeling in this simplified analysis framework, we consider two analysis variants. First, within the analysis window we take $\sigma_i^2 = \sigma^2$ and treat $\sigma$ as a nuisance parameter that we profile over. This allows $\sigma$ to expand beyond $\sqrt{N_i}$, which would result in a weaker limit on the DM lifetime but improved mismodeling, as seen in Fig.~\ref{fig:ts_variants}, whereby this technique dramatically reduces the number of high-TS excesses, though some mismodeling is still present. Lastly, we incorporate the `spurious signal' nuisance parameter method introduced by the ATLAS Collaboration in Ref.~\cite{ATLAS:2014cnc} and later applied to X-ray line searches for decaying DM in Ref.~\cite{Foster:2021ngm}. In this method, a nuisance parameter is introduced in the analysis that is degenerate with the signal parameter except for a Gaussian, zero-mean prior. The variance of the prior $\sigma^2_{{\rm spur},m_\chi}$ at a given mass point $m_\chi$ is determined by the side-bands. In particular, we set 
\es{eq:sigma_spur}{
    \sigma_{\mathrm{spur},m_\chi}^{2} = &\max\!\left[\,0,\ \mathrm{Var}\big[\{\tau_\chi^{-1}\}_{m_\chi}\big]_{\mathrm{observed}} \right. \\ &\left.- \mathrm{Var}\big[\{\tau_\chi^{-1}\}_{m_\chi}\big]_{\mathrm{expected}}\right],
}
where the variances refer to the observed and expected (calculated through MC) variances in the sideband ensemble for that given mass point. For definiteness, we define the sideband at a given mass point $m_\chi$ to be the set of all unmasked test mass points within $\pm 20 \sigma_E$ of $m_\chi$ with $\pm 2 \sigma_E$ also masked. The results of this approach are shown in Fig.~\ref{fig:ts_variants} and Fig.~\ref{fig:cosi_limits} (labeled `energy only: with spurious signal nuis.~param.'). Incorporating the spurious signal nuisance parameter broadly brings the observed distribution of discovery TSs more in line with the expectation of the null hypothesis, at the expense of slightly reduced sensitivity to decaying DM at certain DM masses.

We caution that the mismodeling tests performed above are only performed on 3 months of data, while our sensitivity projections are for 2 years of data. Thus, background mismodeling concerns will be even more important in the real analysis than our mock analysis suggests. On the other hand, in a real analysis our simplified power-law model can likely be improved by including {\it e.g.} a more physics-informed background model. 

\subsection{CDS Analysis}
\label{sec:cds}

As noted in Sec.~\ref{sec:cosi}, the CDS is the data space in which COSI records every event, spanned by four variables: the total energy $E$, the Compton scatter angle ($\phi$), and the direction of the scattered photon which defines the Compton cone axis ($\ell$, $b$). The energy-only analysis of Sec.~\ref{sec:energy_only} is a projection of this space onto its energy axis, obtained by summing over the three angular coordinates. Here we instead work in the full CDS, retaining the spatial information that the
projection discards. Because the signal and the background populate the spatial coordinates differently, retaining them allows the two to be separated in ways the energy spectrum alone cannot. The scatter angle $\phi$ is distributed differently for a narrow line signal within COSI's energy band than for the background, which contains a substantial population of high energy events that are not fully contained and therefore incorrectly reconstructed,\footnote{The reconstructed $\phi$ follows from applying the Compton formula to the measured energy depositions. For a $\gamma$-ray that is fully absorbed in the detector the true scattering angle can be determined, and the distribution of $\phi$ follows the Klein--Nishina cross section. When an event is not fully contained, the total measured energy underestimates the incident energy and the reconstructed $\phi$ no longer corresponds to a physical scattering angle, so these events populate $\phi$ differently.} while ($\ell$, $b$) encode the sky morphology of the emission.

To start, we bin the data in the binning described in Sec.~\ref{sec:sig_bkg}. However, unlike the DM signal for which we generate 100 MC realizations at each mass, current computing limitations constrain us to only a single realization of the  background. As a result, once binned into the full CDS, the counts per bin become notably low, the empty background bins dominate the likelihood, and Wilks' theorem does not apply. We address this issue by background smoothing.

Smoothing the background is possible because, even with only one background realization, the spatial shape of the angular components in the continuum does not differ greatly across suitably chosen energy bins. We therefore average the background model in spatial analysis bins across the energy coordinate within a sliding energy window.  Bright lines are excluded from the sliding-window average. Instead, the bins within $\pm 5 \sigma_E$ of each bright line are averaged only among themselves. The sliding window sizes are energy dependent and described in App.~\ref{app:cds_smoothing}. Once smoothed, we perform a toy MC and confirm that the TS follows the asymptotic $\chi^2_1$ distribution and that the $\mathrm{TS}$ distribution gives the correct 95\% one-sided coverage at the $\mathrm{TS} \approx 2.71$ threshold. 

Given that the expected number of counts in many analysis bins are less than one, we use an analytic continuation of the Poisson likelihood for our Asimov-based projections,
\begin{equation}
    \ln \mathcal{L}(A) = \sum_i \left[ n_i \ln \mu_i(A) - \mu_i(A) - \ln \Gamma(n_i + 1) \right],
    \label{eq:cds_loglikelihood}
\end{equation}
where $n_i$ is the number of counts in CDS bin $i$, $\mu_i(A) = A\,s_i + b_i$ is the number of expected counts, with $A$ a proxy for the DM-decay signal strength, and $b_i$ the smoothed background. We follow the same construction as the optimal energy-only analysis, computing the $95\%$ confidence one-sided expected limit on the Asimov data set with no energy window. In particular, to compute expected upper limits we set $n_i = b_i$, giving the best-fit signal strength $\hat{A} = 0$, and the upper-limit TS reduces to
\begin{equation}
    \mathrm{TS}(A) = 2 \sum_i \left[ A s_i - b_i \ln \left( 1 + \frac{A s_i}{b_i} \right) \right] \,.
    \label{eq:cds_ts_asimov}
\end{equation}

The result, shown in Fig.~\ref{fig:cosi_limits} as `optimal CDS analysis', is an \emph{optimistic} bound: it isolates the best case available from the spatial and spectral information, which a realistic CDS analysis carrying background nuisance parameters would not fully achieve. The optimal CDS analysis provides a minor improvement over the optimal energy-only analysis with a median factor of $\sim1.15$ (see App.~\ref{app:energy_only_v_CDS}).

\section{Results}
\label{sec:results}

The main results of this work are the projected $95\%$ confidence lower limits on the DM decay lifetime into two photons, $\tau_\chi$, as a function of the DM mass $m_\chi$ (see Fig.~\ref{fig:cosi_limits}). For comparison, in gray we include the current leading constraints in the MeV band from INTEGRAL/SPI~\cite{Calore:2022pks,Fischer:2022pse} and COMPTEL~\cite{Essig:2013goa}. We find that across our main analyses COSI is broadly projected to reach lifetimes $\tau_\chi \sim \mathcal{O}(10^{30})$~s below $m_{\chi} \sim 4$~MeV, improving upon the strongest existing limits by over an order of magnitude. The expected sensitivity weakens above a few MeV primarily due to COSI's decreased effective area, which comes from worse containment of higher energy $\gamma$-rays. We also illustrate the comparison to the initial point-source sensitivity-based projections for COSI inferred in Ref.~\cite{Caputo:2022dkz}. Incorporating the full COSI instrument response and applying it to the extended DM source in this work, we find that our improvement in sensitivity primarily comes from utilizing the entire sky throughout COSI's projected 2 year mission, as opposed to the, \textit{e.g.}, Galactic Center source assumed in Ref.~\cite{Caputo:2022dkz}. 

\begin{figure}[t]
   \centering
   \includegraphics[width=0.49\textwidth]{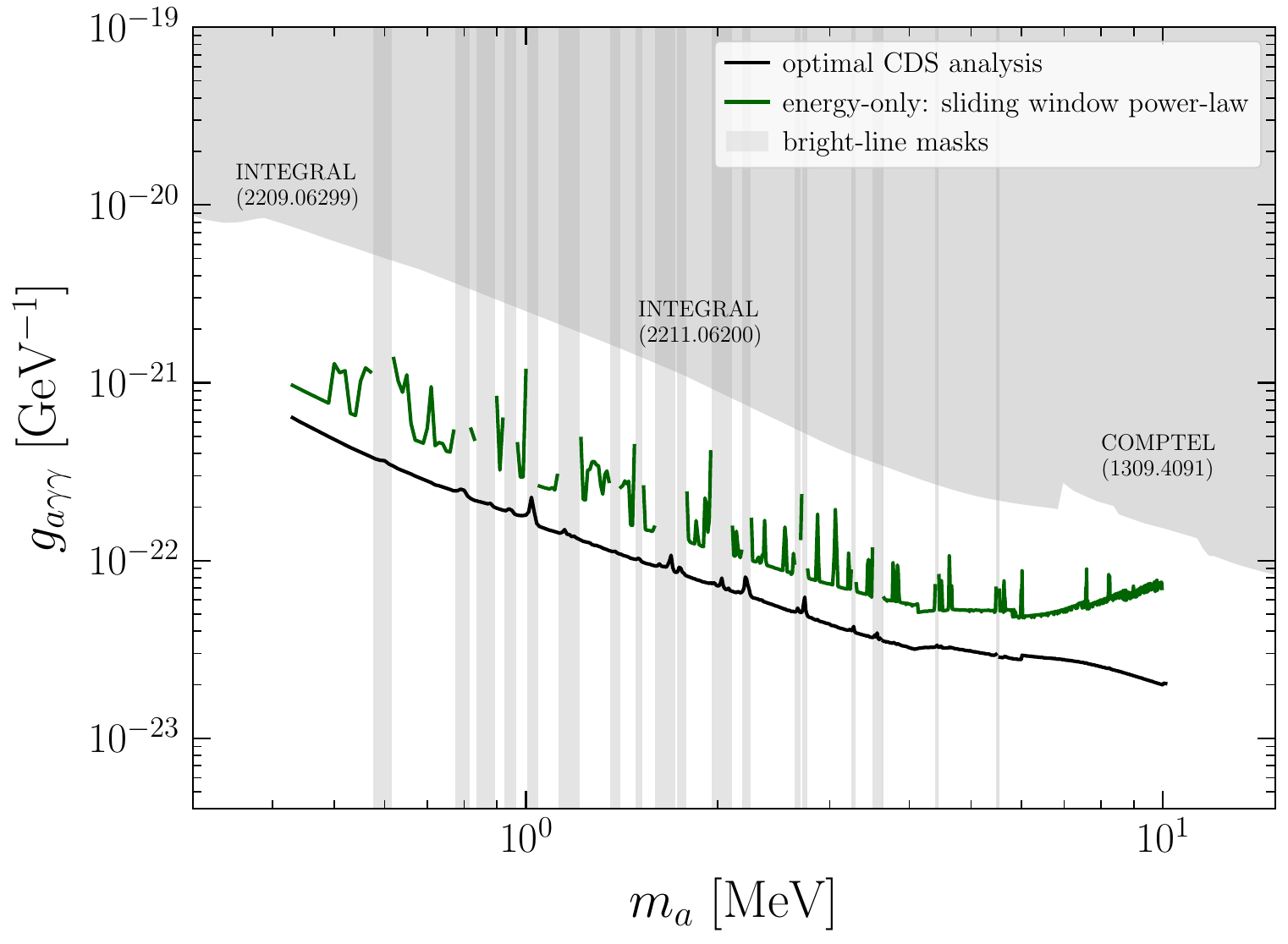}
   \caption{Projected COSI constraints on the DM decay-lifetime limit recast as the ALP--photon coupling $g_{a\gamma\gamma}$, showing the expected 95\% confidence constraints from our energy-only with a sliding window power-law spectral analysis as well as from our CDS analysis. We compare against recast limits from Refs.~\cite{Calore:2022pks,Fischer:2022pse,Essig:2013goa}.}
   \label{fig:axion_lim}
\end{figure}

We may translate our lifetime constraints into projected limits on specific particle DM scenarios. As an example, we consider axion-like particles in the MeV mass range with axion-photon couplings $g_{a\gamma\gamma}$ (see Refs.~\cite{Marsh:2015xka,Hook:2018dlk,Safdi:2022xkm,Caputo:2024oqc} for reviews). The partial decay rate for $a \to \gamma \gamma$ is
\begin{equation}
    \Gamma(a \to \gamma \gamma) = \frac{g_{a\gamma\gamma}^2\, m_a^3}{64\, \pi} \,.
\end{equation}
Recasting our lifetime constraints accordingly gives the projected sensitivity to $|g_{a\gamma\gamma}|$ shown in Fig.~\ref{fig:axion_lim}.  Note that $|g_{a\gamma\gamma}| \sim \alpha_{\rm EM} / (2 \pi f_a)$, with $\alpha_{\rm EM}$ the fine structure constant, which implies that we expect $|g_{a\gamma\gamma}| \sim 10^{-20}$ GeV$^{-1}$ for typical values $f_a \sim 10^{17}$~GeV of the decay constant that can be expected in string theory compactifications~\cite{Svrcek:2006yi,Demirtas:2018akl,Benabou:2025kgx,Benabou:2026jtv}. Moreover, in some string theory constructions---such as in Type IIB string theory---the axion-photon couplings can be additionally suppressed by factors $\epsilon \sim 10^{-3} - 10^{-1}$ from kinetic mixing~\cite{Gendler:2023kjt}. Considering that many string compactifications predict a nearly log-flat distribution of axion masses~\cite{Arvanitaki:2009fg,Mehta:2021pwf}, including potentially in the MeV mass range, we conclude that the future COSI mission will probe a meaningful parameter space that naturally arises in top-down string theory constructions.

For sterile neutrino DM of mass $m_s$ that decays via $\chi\to\nu\gamma$ with active--sterile mixing angle $\theta$, the partial decay rate to the photon final state of a Majorana sterile neutrino is~\cite{Pal:1981rm,Shrock:1982sc}
\begin{equation}
\begin{split}
\Gamma(\chi\to\nu\gamma)
&=\frac{9\,\alpha_{\rm EM}\,G_F^{2}}{1024\,\pi^{4}}\,
  \sin^{2}(2\theta)\,m_s^{5}\\[2pt]
&= 1.361\times10^{-29}~{\rm s}^{-1}
  \left(\frac{\sin^{2}(2\theta)}{10^{-22}}\right)\!
  \left(\frac{m_s}{1~{\rm MeV}}\right)^{\!5},
\end{split}
\label{eq:sterile_decay_rate}
\end{equation}
where $G_F$ is the Fermi constant. The lifetime constraints map directly onto bounds on $\sin^2(2\theta)$ at fixed $m_s$, with our projected sensitivity reaching $\sin^2(2\theta)$ at and below the level of $10^{-22}$. Mixing angles this small cannot generate the observed relic abundance through active--sterile oscillations~\cite{Dodelson:1993je}, which at $m_s \sim 1$~MeV would require $\sin^2(2\theta) \sim 10^{-13}$; the sterile neutrinos probed here must instead be produced by a mechanism independent of $\theta$, such as decays of a gauge-singlet scalar~\cite{Shaposhnikov:2006xi,Kusenko:2006rh,Petraki:2007gq}, gauge interactions in extended models~\cite{Kusenko:2010ik,Fujisawa:2025yqi}, or oscillations followed by entropy dilution~\cite{Asaka:2006ek}.  We remark, however, that the sterile neutrino lifetime is dominated not by this loop-level two-body decay, but rather by the decay to three active neutrinos and, above 1 MeV, the decay into an electron-positron pair and an active neutrino.

\section{Discussion}
\label{sec:discussion}

We demonstrate COSI's projected sensitivity to decaying DM into two monochromatic photons across the 0.2 to 5~MeV energy band using energy-only spectral analyses and a four-dimensional CDS analysis. For the CDS, we treat the background as perfectly known, projecting a best-case lower limit with no systematics modeling. For the energy-only analyses, we present several methods for mitigating background mismodeling which, beyond the factor of two or more in sensitivity lost in going from the optimal to the sliding-window analysis, have only a minor additional effect on the sensitivity to the DM signal. The energy-only analysis is our main result, projecting COSI to probe DM lifetimes up to roughly $\tau_{\chi} \sim \mathcal{O}(10^{30})$~s, over an order of magnitude improvement over the current limit in Ref.~\cite{Fischer:2022pse} (and up to two orders of magnitude compared with Ref.~\cite{Calore:2022pks}).

Comparing the optimal energy-only and CDS analyses, the gain from the full CDS analysis is marginal (median $\sim$$1.15$, see App.~\ref{app:energy_only_v_CDS}). Since a diffuse NFW halo is not sharply distinguished from the diffuse background by morphology, this marginal improvement might be expected. We note that the energy-only analysis is well suited to scanning the full mass range and handling mismodeling effects, while the CDS analysis would be needed to establish the origin of a candidate line.

Our improvement over past limit projections for COSI (\textit{e.g.}, Ref.~\cite{Caputo:2022dkz}) arises from two main differences. First, Ref.~\cite{Caputo:2022dkz} considered the $3\sigma$ narrow line sensitivity for point sources reported in Ref.~\cite{2022icrc.confE.652T} whereas we computed a 95\% one-sided exclusion limit by forward modeling the DM signal through an updated detector response. Second, Ref.~\cite{Caputo:2022dkz} scaled that point-source sensitivity to a $10^\circ$ region around the Galactic Center following the heuristic prescription of Ref.~\cite{2022ApJ...927..225N} and using the required angular resolution from Ref.~\cite{2022icrc.confE.652T}. Our analysis instead forward models an extended source and collects signals from the full halo, increasing our exposure time and thus our limit. We show the effects of restricting to different regions of interest in App.~\ref{app:ROI}.

In addition to extending bounds for DM decays that create narrow lines, there are several other notable DM channels that emit broad continuum signals in the MeV range, such as DM decay to unstable particles or three-body final states~\cite{Essig:2013goa,Coogan:2019qpu,ODonnell:2024aaw} and Hawking radiation from primordial black holes~\cite{Carr:2020gox,Coogan:2020tuf,Bird:2022wvk,Berteaud:2022tws}. DM annihilation~\cite{Boddy:2015efa, Cirelli:2020bpc, ODonnell:2024aaw} can also produce line-like signals, such as direct annihilation to photons, while annihilation to other final states generally yields a continuum. Searching for continuum signatures requires a fundamentally different statistical treatment. We plan to assess the sensitivity of COSI to these scenarios in future work.

Projecting the sensitivities in this paper required simulating the DM signals and the backgrounds directly using MEGAlib, a computationally intensive MC process. However, this workflow is expected to become significantly more accessible prior to COSI's launch with the final release of \textit{cosipy}~\cite{martinez_2023_cosipy}, a Python-based analysis framework currently under development by the COSI Collaboration. A core component of \textit{cosipy} is a source injector module designed to allow users to define arbitrary spatial and spectral models and forward-model them through a pre-computed detector response. This will replace having to directly simulate sources in MEGAlib and allow for direct analysis of COSI flight data once on orbit. Many \textit{cosipy} tools are currently available; however, the software was not used for most of this study because the existing detector response only supports a binning that is too coarse for narrow-line analyses such as those in this work. These limitations are expected to be addressed in the final release of \textit{cosipy}. 

COSI's launch in 2027 will be a major step forward for MeV astrophysics and beyond the Standard Model physics, opening the possibility to greatly improved DM decay-lifetime sensitivity in the MeV gap and potentially leading to the discovery of the identity of DM.


\section*{Acknowledgments}

The Compton Spectrometer and Imager is a NASA Explorer project led by the University of California, Berkeley with funding from NASA under contract 80GSFC21C0059. Resources supporting this work were provided by the NASA High-End Computing (HEC) Program through the NASA Advanced Supercomputing (NAS) Division at Ames Research Center. O.N. and B.R.S. are supported in part by the DOE award DE-SC0025293. B.R.S. acknowledges support from the Alfred P. Sloan Foundation and hospitality from the CERN theory group, where much of this work was completed. The work of O.N. is supported in part by the NSF Graduate Research Fellowship Program under Grant DGE2146752. M.D. is supported through a cooperative agreement with the Center for Research and Exploration in Space Sciences and Technology II (CRESST II) between NASA Goddard Space Flight Center and University of Maryland, College Park, under award number 80GSFC24M0006. S.M. and T.T. are supported by the Grant-in-Aid for Scientific Research (A), Grant No.24H00244. S.M. is also supported by the Grant-in-Aid for Transformative Research Areas (A), Grant No. 24H02244. H.Y. is supported by JSPS KAKENHI Grant Number 23K13136, 26H00833. This research used resources of the National Energy Research Scientific Computing Center (NERSC), a U.S. Department of Energy User Facility, using NERSC award no. HEP-ERCAP0037471.

\bibliography{refs}

\clearpage
\onecolumngrid

\appendix

\section{COSI Event Detection and Simulation Pipeline}
\label{app:cosi_dec_sims_pipeline}

\begin{figure}[htbp]
   \centering
   \includegraphics[width=0.95\textwidth]{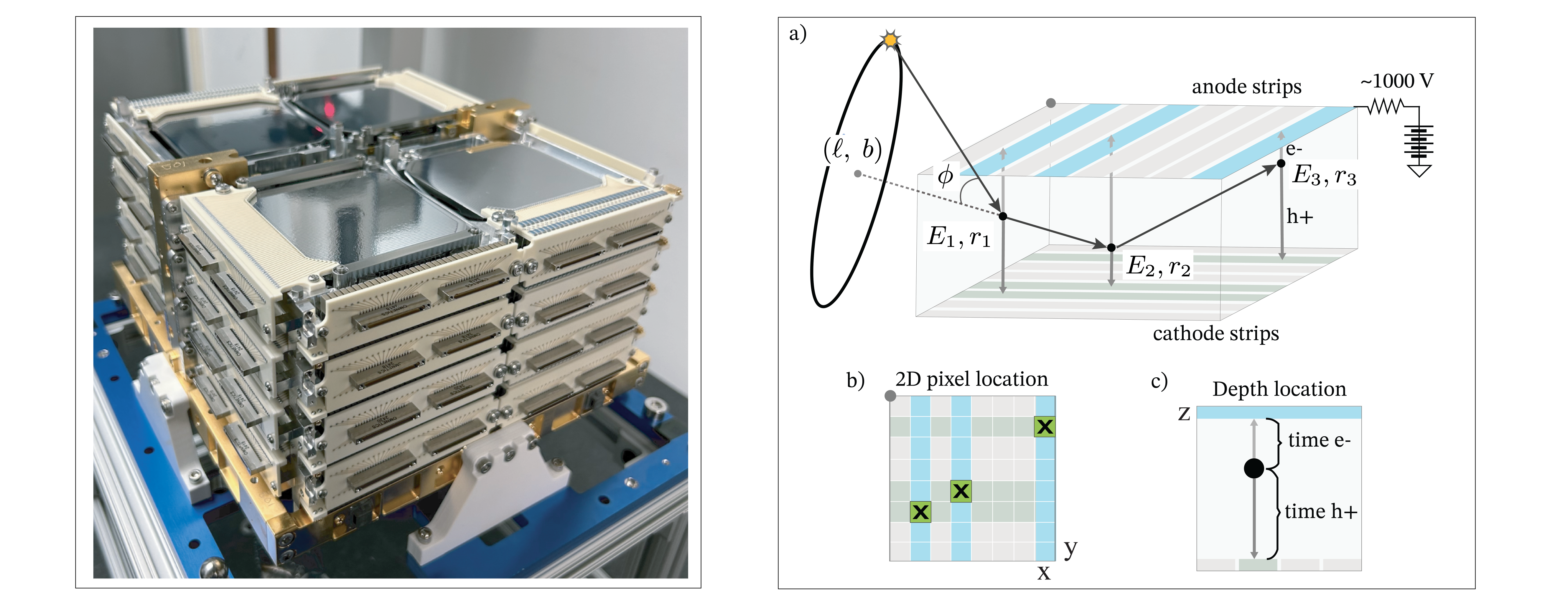}
   \caption{\emph{Left:} Photo of COSI's 16 GeD stack. Each detector face is segmented into 64 strips. \emph{Right:} Illustration of how COSI's GeDs detect events. In a), a $\gamma$-ray enters a detector, scattering twice (($E_1$, $r_1$), ($E_2$, $r_2$)) before being photoabsorbed ($E_3$, $r_3$). At each interaction site a cloud of $e^-$ and $h^+$ pairs are drifted by a ${\sim}1000$~V bias to opposite faces of the detector (this process is denoted by the vertical arrows). The current induced by the $e^-/h^+$ is measured on each strip. The energy and the first two interaction positions, $r_1$ and $r_2$, are then used to calculate the Compton scatter angle $\phi$, constraining the $\gamma$-ray's origin to a Compton cone on the sky with center at ($\ell$, $b$). b) shows how the cross-strips are used to make a pixel grid which determines the ($x$, $y$) position of the interaction, while c) illustrates how the time difference between when the $e^-$ and $h^+$ signals are read out is used to determine the depth $z$ at which the interaction took place.} 
\label{fig:cds_dec}
\end{figure}

COSI's Compton camera consists of sixteen $8\times8\times1.5$ (cm$^3$) GeDs, arranged in a $2 \times 2 \times 4$ array, see Fig.~\ref{fig:cds_dec}. Each GeD face is segmented into 64 thin conducting electrode strips that are read out by custom ASICs (Application-Specific Integrated Circuits)~\cite{2020NIMPA.95461230W} (also see Ref.~\cite{Sleator_inprep}). As a $\gamma$-ray passes through the GeDs, it preferentially Compton scatters before being photoabsorbed. The subsequently ejected electrons move microscopically in the solid, further ionizing the surrounding material and creating a cloud of electron-hole ($e^-/h^+$) pairs. A large voltage bias, $\sim 1000$~V, drifts the $e^-$ and $h^+$ to opposite sides of each GeD where they are incident on the strips~\cite{boggs_2023}, generating a current proportional to the deposited charge, see Fig.~\ref{fig:cds_dec}.a. The total measured energy for an event is obtained by summing the individual energies read out by the strips. To measure the position of the interaction, the strips on opposite GeD faces are orthogonal and their overlap defines the $x, y$ location, see Fig.~\ref{fig:cds_dec}.b. The $z$ position is determined from the difference in collection time between the $e^-$ signal and the $h^+$ signal~\cite{2026NIMPA108671332R}, see Fig.~\ref{fig:cds_dec}.c. The $\gamma$-ray's origin is constrained by using the first two interactions to calculate the Compton scatter angle, $\phi$, which defines the Compton cone with center at ($\ell$, $b$). To reject $\gamma$-rays and charged particles originating in Earth's atmosphere, and partially-absorbed events (\textit{i.e.} incident $\gamma$-rays that Compton scatter and exit the GeDs before being photoabsorbed), the GeD stack is surrounded on five sides by an anti-coincidence shield of bismuth germanate (BGO) scintillators which veto events that pass through it~\cite{2025ExA....60....9C,patel_2025_icrc}. This design reduces backgrounds by both enabling event rejection and reducing the amount of passive material between the GeDs and sources of interest, lowering the instrumental backgrounds. 

Each of the processes described above---charge cloud creation and transport effects~\cite{2025ExA....59....8P,sleator_2019,2022HEAD...1910818B}, strip readout, the timing measurement that sets the depth, and the shield veto---have to be accounted for when forward modeling our DM signal into a projected measurement. Since COSI's forward modeling cannot be described analytically, we must use well-benchmarked MC simulations. This is achieved using MEGAlib~\cite{2006NewAR..50..629Z}.

To simulate our DM signal, we begin by converting the continuous flux from Eq.~\ref{eq:signal_flux} and morphology from Eq.~\ref{eq:NFW} into an all-sky discrete spatial$\,\times\,$spectral flux map. For the spatial component, we evaluate the $\mathcal{D}$-factor in polar coordinates on a $0.5^\circ$ grid. For the spectral component, the line is modeled by a narrow three-point array centered on $E_{\rm peak} = m_\chi/2$, with a  $1$~eV FWHM, which is normalized to ensure the integral is $N_\gamma = 2$. This signal is then scaled by $(4\pi m_\chi \tau_\chi)^{-1}$, with $m_\chi = [0.4 -10]$~MeV and $\tau_\chi = 10^{28}$~s, and passed into MEGAlib's MC simulation software, Cosima~\cite{zoglauer_2009_cosima}. Cosima then samples from the discretized flux map and tracks each photon's path through a detailed 3D model of the COSI instrument. The spacecraft's simulated pointing history is also fed in to update the instrument orientation and livetime over the simulated three-month period. Cosima is built on ROOT and GEANT4; for these simulations, we use ROOT~\cite{1997NIMPA.389...81B} version 6.36.12~\cite{root_6_36} and GEANT4~\cite{2003NIMPA.506..250A,2016NIMPA.835..186A} version 11.2.2.

The output of Cosima is then forward modeled through MEGAlib's detector effects engine, Revan,\footnote{While these simulations employ Revan's detector effect engine, the COSI Collaboration is developing a dedicated package, Nuclearizer, to process flight data into high-level data. Nuclearizer was originally built for the 2016 COSI Balloon Campaign~\cite{2016int..workE..75K}, as detailed in Ref.~\cite{2019HEAD...1710963S} and Ref.~\cite{sleator_2019}, and is now being updated for the COSI SMEX mission. These updates include an instrument-specific detector effects engine to enhance the overall analysis pipeline as outlined in Ref.~\cite{10.1117/12.3103024}. A future analysis that follows these simulation steps should employ Nuclearizer between Cosima and Revan to achieve more accurate results.} which translates the simulation's truth values into realistic detector readouts with noise and uncertainty. Revan also reconstructs the order of the successive scatters in each event, which COSI cannot measure directly~\cite{2007nssm....6...96Z,10.1117/12.3063687}; we use the same trained Bayesian Compton-sequence reconstruction as the COSI Collaboration's background simulations. Each event is saved as a list of individual interactions parameterized by the measured energy deposition and position, $(E_1, x_1, y_1, z_1, \dots)$. From Revan, the events are passed through MEGAlib's Mimrec package, which applies the standard selection cuts: at least one reconstructed Compton scatter, a minimum interaction separation of $1.5$~mm, and event energies must be greater than $0.1$~MeV. We also filter out events with reconstructed energy at exactly $1.8$~MeV, which are an artifact of a hard-coded threshold in Cosima's instrument-model.
 
As the last step, \texttt{cosipy}~\cite{martinez_2023_cosipy} is used to apply the instrument orientation file and transform the event data from MEGAlib, which is saved in the detector's coordinate system, into the Galactic coordinates $(E, \phi, \ell, b)$ that span the CDS. Finally, the instrumental backgrounds are much higher when the spacecraft is passing through the SAA. Therefore, using \texttt{cosipy}, we remove all the intervals where COSI is passing through the SAA. This is done for both the backgrounds and the DM signals. 

\section{Background line catalog}
\label{app:bkg_lines}

\begin{figure}[htbp]
    \centering
    \includegraphics[width=0.95\textwidth]{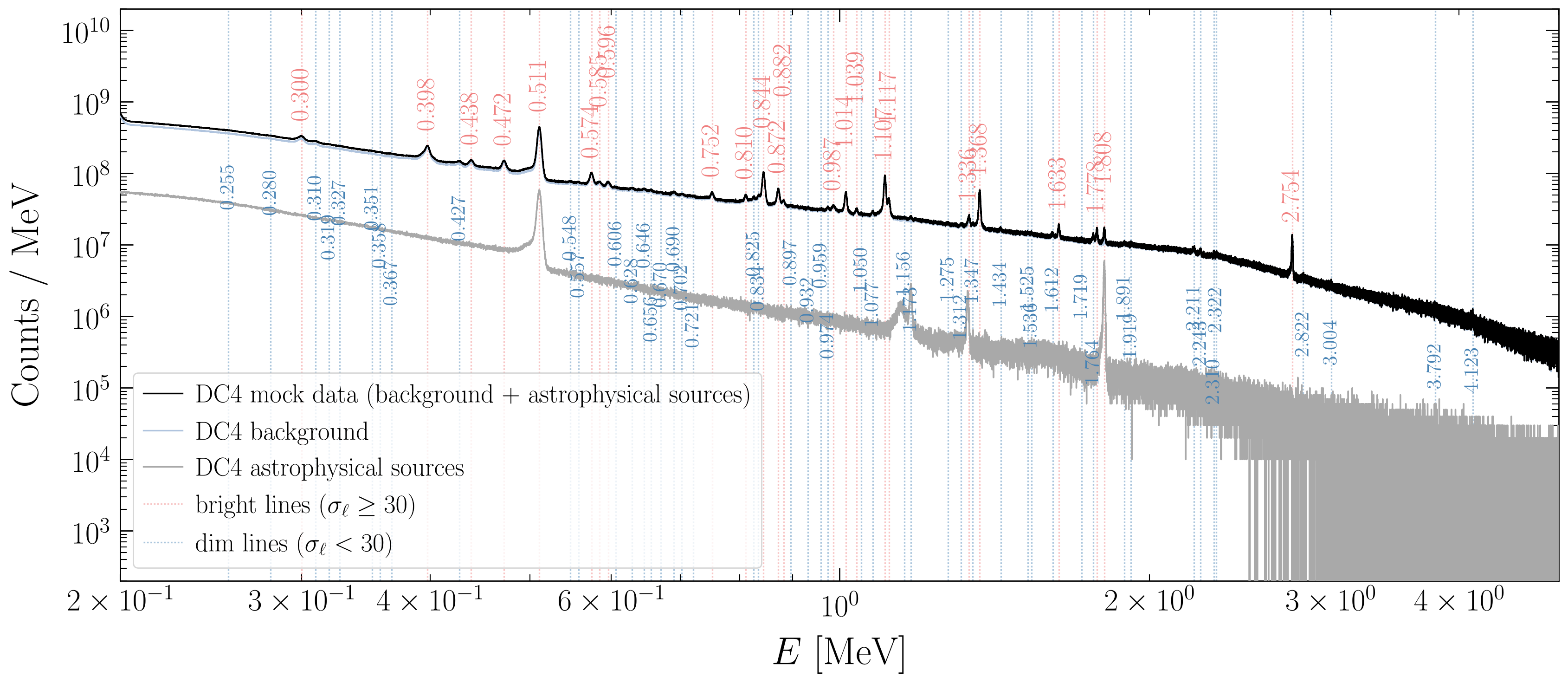}
    \caption{The simulated energy spectra binned in 0.1 keV bins for the DC4 background data (solid light blue) and astrophysical sources (solid gray), which combined form the DC4 mock data (solid back line). Lines are identified through the iterative sliding window analysis discussed in the text. \emph{Dim} lines, defined as those with significance $\sigma_{\ell} < 30$, are indicated in blue. \emph{Bright} lines, defined as those with significance $\sigma_{\ell} \geq 30$, are indicated in red. See Table~\ref{tab:background_lines} for a complete list of lines and their $\sigma_{\ell}$ values.}
    \label{fig:background_lines}
\end{figure}

The background contains many $\gamma$-ray lines which can be mistaken for a DM line, and hence needs specific handling in our analysis. Both Ref.~\cite{2026ApJ...997..284G} (see Tables 1-3) and Ref.~\cite{2025arXiv250810172V} have studied COSI's expected background lines. However, there is not yet a complete catalog of background lines and their expected brightness. Therefore, for our purposes, we build a list of lines that we use to determine which lines to mask and which to model in our analysis. 

To compile this list, we identify background lines by adopting our sliding window analysis introduced in the main text, and scanning the window across the DC4 mock data spectrum, rebinned in intervals of 0.1 keV to precisely determine line locations. In each window, we fit the continuum with a local power-law plus a Crystal Ball line profile to assign candidate lines their discovery TS, and the energy corresponding to the peak of the TSs across windows marks a line's location. We first identify the brightest lines and add them to the background model. The scan is continuously repeated so that successively weaker lines are found in order of decreasing significance, until no line above a significance threshold of $\sigma_{\ell} = \sqrt{\rm TS} = 3$ remains. This yields a data-driven catalog of line energies and strengths, which are recorded in Table.~\ref{tab:background_lines}. Fig.~\ref{fig:background_lines} marks the identified lines on the background spectrum. In this work, lines with $\sigma_{\ell} \geq 30$ are categorized as bright lines and the rest are categorized as dim lines.

\begin{table*}[htbp]
\centering
\caption{Identified $\gamma$-ray lines in the DC4 mock data and their significances $\sigma_{\ell}$.}
\label{tab:background_lines}
\small
\vspace{0.2cm}
\begin{tabular}{c c|c c|c c|c c|c c|c c}
\toprule
Line (keV) & $\sigma_{\ell}$ & Line (keV) & $\sigma_{\ell}$ & Line (keV) & $\sigma_{\ell}$ & Line (keV) & $\sigma_{\ell}$ & Line (keV) & $\sigma_{\ell}$ & Line (keV) & $\sigma_{\ell}$ \\
\midrule
254.7 & 7.6 & 472.1 & 155.0 & 690.4 & 27.3 & 959.0 & 7.4 & 1312.0 & 10.4 & 1808.3 & 88.8 \\
279.9 & 8.3 & 510.8 & 983.5 & 702.5 & 7.6 & 974.2 & 9.8 & 1336.2 & 65.1 & 1891.0 & 10.2 \\
300.1 & 100.1 & 547.5 & 8.2 & 720.5 & 7.5 & 986.6 & 40.9 & 1347.0 & 7.8 & 1920.6 & 7.7 \\
309.6 & 29.0 & 557.5 & 9.1 & 751.9 & 67.5 & 1014.0 & 188.8 & 1368.3 & 313.4 & 2210.9 & 21.1 \\
319.1 & 5.4 & 574.0 & 140.1 & 810.4 & 56.1 & 1039.2 & 43.0 & 1434.0 & 17.0 & 2243.2 & 20.3 \\
326.8 & 6.5 & 584.6 & 36.1 & 825.2 & 8.3 & 1050.2 & 5.0 & 1524.5 & 5.4 & 2309.8 & 7.2 \\
351.4 & 8.5 & 595.9 & 49.6 & 833.9 & 29.1 & 1076.8 & 23.5 & 1535.8 & 5.2 & 2322.2 & 9.0 \\
357.6 & 4.9 & 606.1 & 9.7 & 843.5 & 345.8 & 1106.8 & 345.2 & 1611.5 & 19.8 & 2753.8 & 184.4 \\
366.9 & 6.6 & 628.4 & 15.6 & 871.9 & 150.5 & 1117.0 & 139.7 & 1633.2 & 76.7 & 2821.1 & 10.3 \\
397.8 & 256.1 & 645.8 & 15.0 & 882.5 & 32.6 & 1155.5 & 7.3 & 1718.8 & 9.9 & 3002.3 & 5.3 \\
427.2 & 25.5 & 656.0 & 11.7 & 896.6 & 3.2 & 1173.2 & 18.7 & 1764.2 & 22.0 & 3792.2 & 4.9 \\
438.4 & 86.1 & 670.5 & 5.1 & 932.8 & 5.9 & 1275.2 & 9.1 & 1778.2 & 72.0 & 4122.9 & 10.5 \\
\bottomrule
\end{tabular}
\end{table*}

This list of background lines is used by both our energy-only and CDS analyses. In the energy-only analysis, background lines can mimic a DM line. For dim lines, we can add parameters to model them without having them absorb our signal. Bright lines are instead masked, and we do not report limits at the corresponding masses, since the analysis has little sensitivity there. For the CDS analysis we have to smooth the background to overcome empty cells, and bright lines are smoothed in their own window to preserve their spatial information (see App.~\ref{app:cds_smoothing}).  

We emphasize that this procedure serves only to sort identified lines into bright and dim categories. The origin of a given line is not relevant to the analysis, nor is the absolute accuracy of the fitted significance. What matters is that every line is treated identically, so that the classification is applied consistently across the search band.

\section{The CDS Analysis}
\label{app:cds}

We describe the treatment of the background for the CDS analysis in App.~\ref{app:cds_smoothing}. Next, we compare the CDS limits to the optimal energy-only limits in App.~\ref{app:energy_only_v_CDS}. Lastly, we use the CDS to apply a spatial cut and consider limits on the inner $50^\circ$ around the Galactic Center in App.~\ref{app:ROI}.

\subsection{Smoothing the backgrounds in the CDS}
\label{app:cds_smoothing}

\begin{figure}[htbp]
    \centering
    \includegraphics[width=0.8\textwidth]{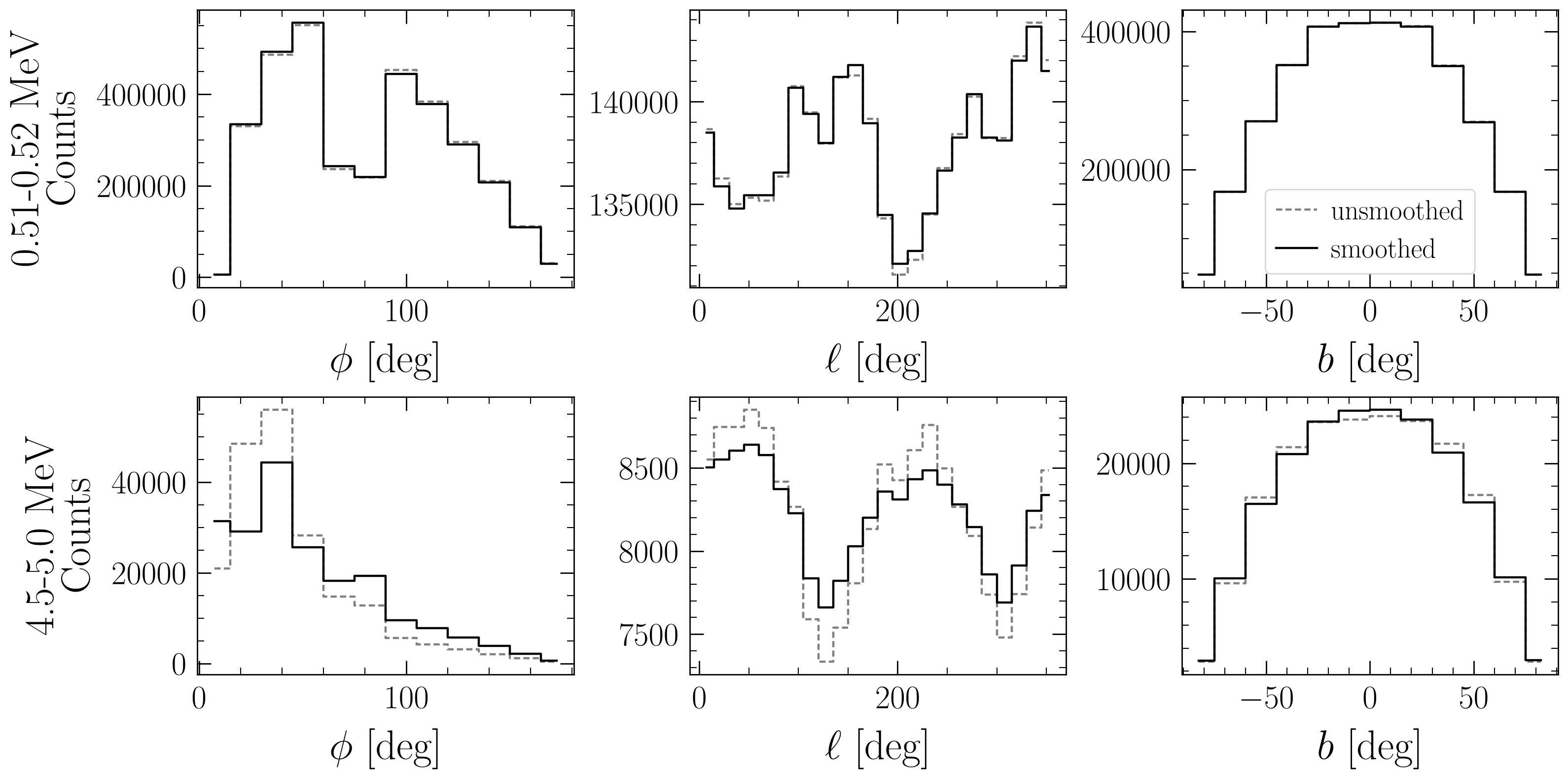}
    \caption{Results of applying block smoothing to the DC4 mock data. The unsmoothed version is shown in dashed gray, while the smoothed version is shown in solid black. \emph{Top:} An example of smoothing the bright 0.511~MeV line, within its $\pm 5 \sigma_E$ window. Because the 0.511~MeV line contains substantial counts, the spatial structure is well-preserved. \emph{Bottom:} An example of smoothing over the 4.5-5.0~MeV continuum block. This block has fewer counts, and we therefore see more distortion in the spatial components.}
    \label{fig:smoothing_marginals}
\end{figure}

As noted in Sec.~\ref{sec:cds}, the mock DC4 data only exists as a single realization. While the counts are high in the energy-only case (summing over the CDS angles), the counts per bin are sparse once binned over all four variables. This results in the TS, $q \equiv \sum_i q_i$, (under the Asimov convention $n_i = b_i$) being dominated by $2 A s_i$ contributions from empty cells:  

\begin{equation}
    q_i = \left\{
    \begin{array}{c@{\qquad}l}
        2\left[A s_i - b_i \ln\left(1 + \dfrac{A s_i}{b_i}\right)\right],
            & b_i > 0, \\[1.2em]
        2 A s_i, & b_i = 0 .
    \end{array}
    \right.
    \label{eq:q_cell_contribution}
\end{equation}

We note that this is only valid when $b_i = 0$ is actually expected from the data. For our background, however, an empty cell is mostly an artifact of having a single MC realization, yet empty cells completely dominate the test statistic and their linearity in the signal $A$ makes the resulting limit artificially stronger. While this could be mitigated by modeling the background, ($E$, $\phi$, $\ell$, $b$) are all convolved and cannot be described by a simple flat, linear, or power law model. We therefore do not pursue a parametric CDS background here and instead implement two alternatives: first, we smooth the background, and second, we explore coarser binning of the spatial variables. To evaluate each, we define $q_{\rm empty}$ as the fraction of the TS coming from cells with $b_i = 0$. As $q_{\rm empty}$ approaches 1, the limit is being dominated by the empty bins. Our goal is to minimize $q_{\rm empty}$ in both approaches. 

To smooth the backgrounds, we utilize the fact that, while we cannot average over independent realizations, we can average over energy within the single simulation we have because the angular distribution of the background continuum changes slowly with energy. To do this, the energy axis is split into blocks, described in Table~\ref{tab:binning_decisions}. Within each block we sum the angular counts within the energy bins over the entire block. From this, we create a probability density that describes the counts in each angular bin. We then redistribute each energy bin's own total counts onto that shared density, which ensures that the energy spectrum is unchanged and that only the spatial shape is averaged across a block. While smoothing, we require that values are only averaged into kinematically allowed Compton scatter angles, $\phi$. Once smoothing is complete, we verify that the total counts in each energy bin remain the same, and that no counts appear in forbidden bins. We quantify the change in the spatial features due to smoothing by determining the variation in the excess over the statistical fluctuation of the spatial components before and after smoothing, which we call the \emph{distortion}.
 
\begin{figure}[htbp]
    \centering
    \includegraphics[width=0.85\textwidth]{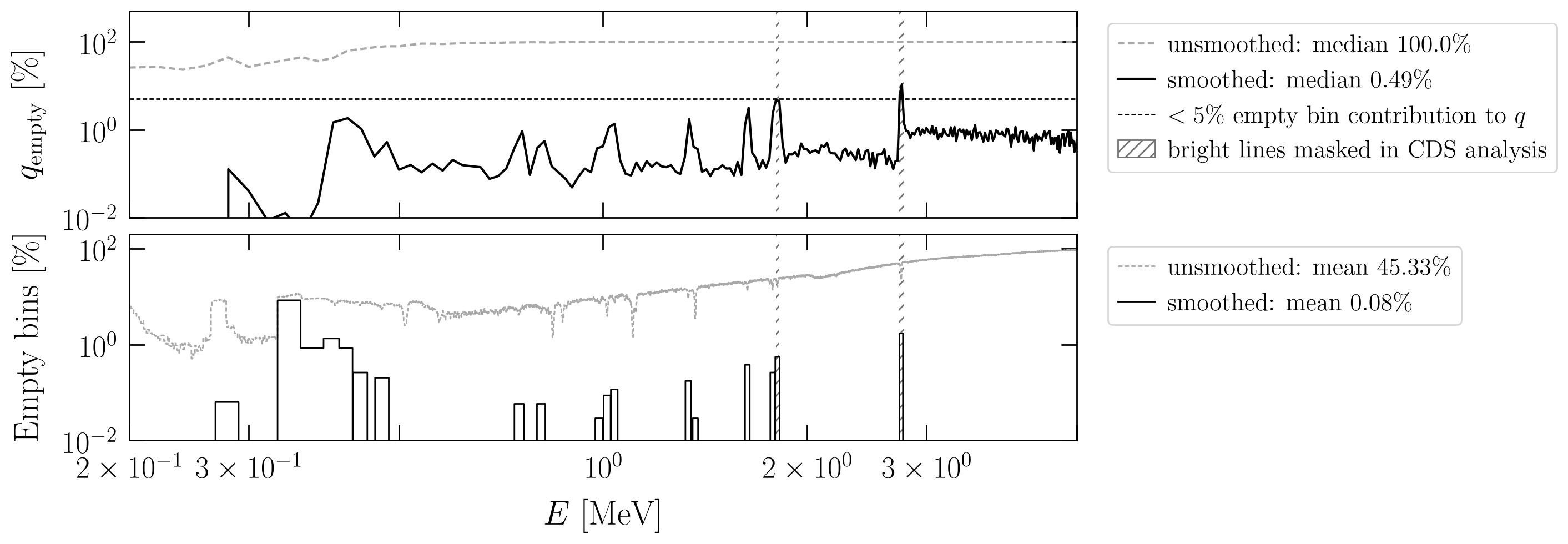}
    \caption{\emph{Top:} The fraction,  $q_{\rm empty}$, of the TS contributed by bins with $b_i =0$, before and after the energy-block averaging. \emph{Bottom:} The fraction of kinematically allowed cells that are empty, as a function of energy. The hatched bands indicate the bright lines which, even after smoothing, still have a $q_{\rm empty}>5\%$ and are therefore masked from the reported CDS limits.}
    \label{fig:smoothing}
\end{figure}

While we argue that the continuum's angular features vary slowly with energy, this is not necessarily true at lines, particularly bright lines, which we can expect to have distinct spatial shapes. Therefore, we do not include bright lines in this averaging. Each bright line is instead averaged over its own $\pm 5\sigma_E$ window. Where a bright line still results in a large $q_{\rm empty}$ value, we mask the corresponding DM mass from the reported limits. Dim lines remain inside the continuum blocks, because they do not have enough counts to be averaged on their own. However, their inclusion in the continuum smoothing does lead to the larger distortions, see Table.~\ref{tab:binning_decisions}. We conclude that smoothing decreases the mean $q_{\rm empty}$ from $100\%$ to $0.49\%$, while the percentage of empty bins drops from $45.33\%$ to $0.08\%$, see Fig.~\ref{fig:smoothing}. 

\begin{table*}[htbp]
\centering
\begin{minipage}[t]{0.40\textwidth}
\centering
\textbf{Continuum + dim lines}\\[0.3em]
\begin{tabular}{l c c c}
\toprule
$E$ range & Block size & $q_{\rm empty}$ & Distortion \\
{}[keV] & [keV] & [\%] & [\%] \\
\midrule
$200$--$300$   & 20   & 0.18 & $0.87$--$1.81$  \\
$300$--$800$   & 50   & 0.34 & $1.18$--$7.64$  \\
$800$--$1500$  & 100  & 1.65 & $0.47$--$3.25$  \\
$1500$--$2000$ & 200  & 1.13 & $0.75$--$2.00$  \\
$2000$--$3000$ & 940  & 0.26 & 12.34           \\
$3000$--$5000$ & 2000 & 1.76 & 11.20           \\
\midrule
All            &      & 0.92 & $0.47$--$12.34$ \\
\bottomrule
\end{tabular}
\end{minipage}
\hspace{0.02\textwidth}
\begin{minipage}[t]{0.55\textwidth}
\centering
\textbf{Bright lines}\\[0.4em]
\begin{tabular}{c c c l|c c c l}
\toprule
$E$ & $q_{\rm empty}$ & Distortion & & $E$ & $q_{\rm empty}$ & Distortion & \\
{}[keV] & [\%] & [\%] & & {}[keV] & [\%] & [\%] & \\
\midrule
300.1 & $0.00$ & $1.39$ & keep & 882.5 & $0.28$ & $1.33$ & keep \\
397.8 & $1.00$ & $1.03$ & keep & 986.6 & $0.32$ & $-0.01$ & keep \\
438.4 & $0.56$ & $0.58$ & keep & 1014.0 & $1.49$ & $0.56$ & keep \\
472.1 & $0.33$ & $1.09$ & keep & 1039.2 & $1.13$ & $-0.08$ & keep \\
510.8 & $0.00$ & $0.52$ & keep & 1106.8 & $0.00$ & $1.26$ & keep \\
574.0 & $0.00$ & $0.41$ & keep & 1117.0 & $0.00$ & $4.05$ & keep \\
584.6 & $0.00$ & $0.93$ & keep & 1336.2 & $2.27$ & $0.13$ & keep \\
595.9 & $0.00$ & $0.94$ & keep & 1368.3 & $0.73$ & $0.12$ & keep \\
751.9 & $0.57$ & $0.17$ & keep & 1633.2 & $4.35$ & $-0.21$ & keep \\
810.4 & $0.46$ & $0.06$ & keep & 1778.2 & $3.31$ & $-0.07$ & keep \\
843.5 & $0.00$ & $0.08$ & keep & 1808.3 & $8.37$ & $0.11$ & mask \\
871.9 & $0.75$ & $0.19$ & keep & 2753.8 & $50.27$ & $0.12$ & mask \\
\bottomrule
\end{tabular}
\end{minipage}
\caption{\emph{Left:} The energy range and corresponding block sizes chosen for the continuum smoothing, along with the resulting $q_{\rm empty}$ and distortion percentages. \emph{Right:} The 24 bright lines, with $q_{\rm empty}$ evaluated at $A = 10^{-2}$ within the $\pm 5\sigma_E$ window. Bright lines where $q_{\rm empty}$ is above $5\%$ are masked from the reported limits.}
\label{tab:binning_decisions}
\end{table*}

While smoothing works at some energies, there are shortcomings at higher energies. The two blocks above 2~MeV distort the spatial information by $11$--$12\%$, an order of magnitude more than any block below them. In addition, bright lines above $\sim0.75$~MeV still have too high a $q_{\rm empty}$ value to be included in the limit. 

As a second approach to computing the CDS limit we instead employ coarser binning. For the smoothed background all the spatial components are binned in $15^\circ$ bins. Here we instead loop over a range of binning schemes to find one that maximizes the number of bins and minimizes the number of empty bins. We find small values of $\phi$ contribute greatly to separating the signal from the background and we therefore choose the bin $\phi$ in non-linear bin sizes to maximize the separation it offers at lower scatter angles. For $\ell$ and $b$, the separation they offer gets sacrificed to ensure there are enough counts in each bin. The final binning choice is presented in Table.~\ref{tab:cds_binning}.

\begin{table}[htbp]
\centering
\begin{minipage}{0.3\textwidth}
\centering
\begin{tabular}{c c}
\toprule
\multicolumn{2}{c}{$\phi$ bins} \\
\midrule
Range & Bin size \\
\midrule
$0$--$20^\circ$   & $20^\circ$  \\
$20$--$40^\circ$  & $20^\circ$  \\
$40$--$60^\circ$  & $20^\circ$  \\
$60$--$180^\circ$ & $120^\circ$ \\
\bottomrule
\end{tabular}
\end{minipage}
\hspace{0.5em}
\begin{minipage}{0.40\textwidth}
\centering
\begin{tabular}{l c c c}
\toprule
\multicolumn{4}{c}{$\ell$ and $b$ bins} \\
\midrule
$E$ range & Bin size & $\ell$ bins & $b$ bins \\
\midrule
$200$--$2000$~keV  & $30^\circ$ & 12 & 6 \\
$2000$--$4000$~keV & $60^\circ$ & 6  & 3 \\
$4000$--$5200$~keV & $90^\circ$ & 4  & 2 \\
\bottomrule
\end{tabular}
\end{minipage}
\caption{CDS spatial binning chosen to maximize the number of bins and minimize the number of empty background bins. \emph{Left:} The binning for $\phi$ bins, which is fixed at all energies. \emph{Right:} the $\ell$ and $b$
binning, which coarsens with energy as the total number of background counts decreases.}
\label{tab:cds_binning}
\end{table}

\begin{figure}[htbp]
    \centering
    \includegraphics[width=1\textwidth]{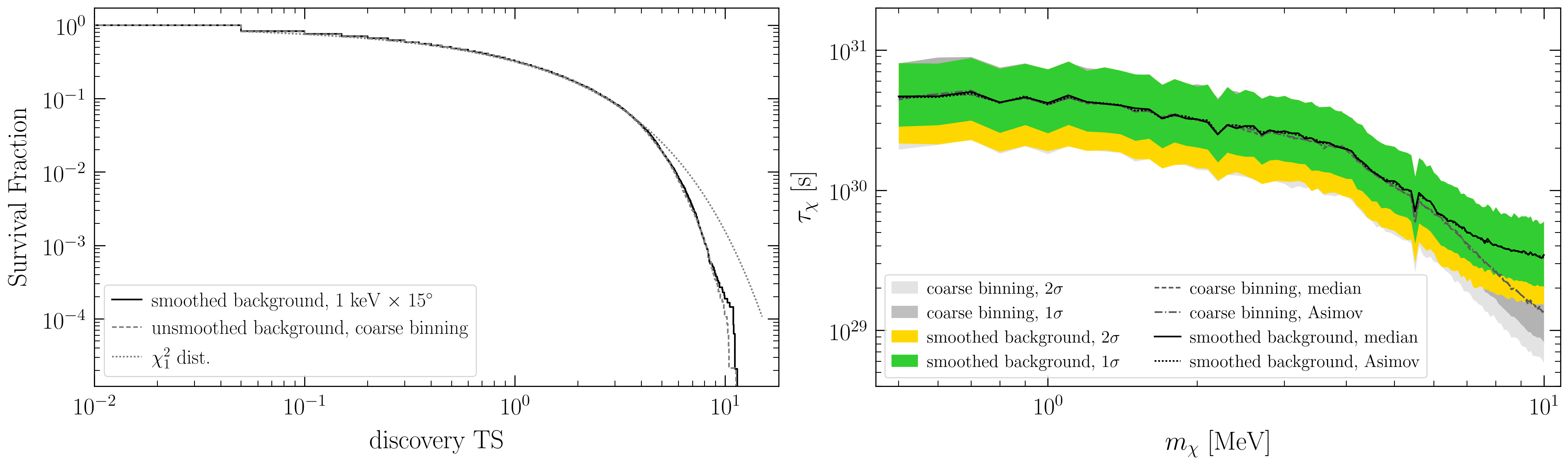}
    \caption{\emph{Left:} The discovery TS survival fraction of the 96 sampled masses, for both the smoothed and the unsmoothed coarse spatial binning background, compared to the $\chi^2_1$ distribution. \emph{Right:} Comparing the lower limit on $\tau_{\chi}$ derived from the Asimov dataset (labeled `Asimov') to the median limit derived from the Poisson-fluctuated datasets (labeled `median'), for both background treatments.}
    \label{fig:survival_and_limits}
\end{figure}

As a last step, for both the smoothed and coarse binning treatments, we verify two properties of our limit-setting procedure. First, that Wilks' theorem applies, so that the test statistic broadly follows the asymptotic $\chi_1^2$ distribution. We verify this by calculating $q$ for 1000 Poisson draws of $n_i = P(b_i + A s_i)$ for DM masses from 0.5 to 10~MeV in steps of 0.1 MeV (96 masses) and 6 different values of $A$. The resulting TS survival fraction is shown in the left panel of Fig.~\ref{fig:survival_and_limits}. Second, we verify that the limit evaluated on the Asimov dataset reproduces the median limit of an ensemble of Poisson-fluctuated datasets. For this, we calculate the limit on $\tau_{\chi}$ for 1000 Poisson draws of $n = P(b_i)$ at each mass, and confirm that the median limit over the draws approaches the Asimov limits, shown in the right panel of Fig.~\ref{fig:survival_and_limits}. We conclude that the counts per bin are large enough for both assumptions to apply, and proceed to calculate projected limits using the Poisson likelihood. 

\subsection{Energy-only vs. CDS}
\label{app:energy_only_v_CDS}

\begin{figure}[htbp]
    \centering
    \includegraphics[width=0.9\textwidth]{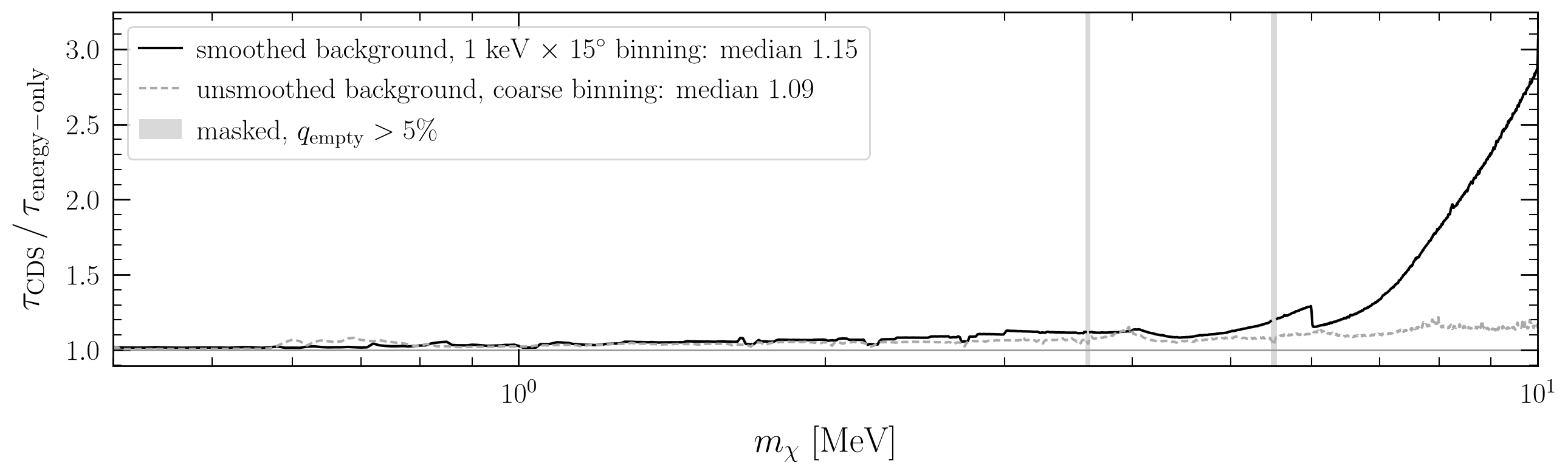}
    \caption{Ratio of the two CDS limits to the optimal energy-only limit. We note that no energy window is applied to any of the limits. The smoothed CDS limit (solid black) shows the greater improvement, while the CDS limit with coarser binning (dashed gray) stays close to the optimal energy-only case (solid gray line at unity).}
    \label{fig:energy_only_vs_CDS}
\end{figure}

For an analysis using flight data, the energy-only and CDS analyses are complementary to each other in setting limits. Much of this work is devoted to background mismodeling, which we expect to be the dominant systematic uncertainty, and it is precisely this systematic uncertainty that would make line searches in the CDS difficult. Therefore, one approach would be to scan the mass range for notable narrow lines using the energy-only analysis first, taking into account systematic mismodeling as is done in this work. This could then be accompanied with a CDS analysis; if a target line was identified, the CDS would be required to separate a real halo signal from, \textit{e.g.}, an unmodeled instrumental line. 

\subsection{Regions of interest in the CDS}
\label{app:ROI}
COSI constrains individual events to Compton cones on the sky, making spatial cuts possible. Here, we consider a $50^\circ$ region around the Galactic Center, chosen to be comparable to INTEGRAL/SPI's $\pm 47.5^\circ$ region studied in Ref.~\cite{Calore:2022pks}. While the CDS does enable spatial regions to be studied, it is not a straightforward cut on $\ell$ and $b$. An event is constrained to the projection of the Compton cone base onto the sky, called a Compton circle, with radius set by $\phi$. Therefore, the $\gamma$-ray's origin is not measured as a precise location, but rather has some probability of being anywhere on the Compton circle, with the width of the Compton circle being set by the angular resolution of COSI (the requirement is ${\sim}2.1^\circ$ FWHM at $1.8$~MeV). 

\begin{figure*}[htbp]
   \centering
   \includegraphics[width=1\textwidth]{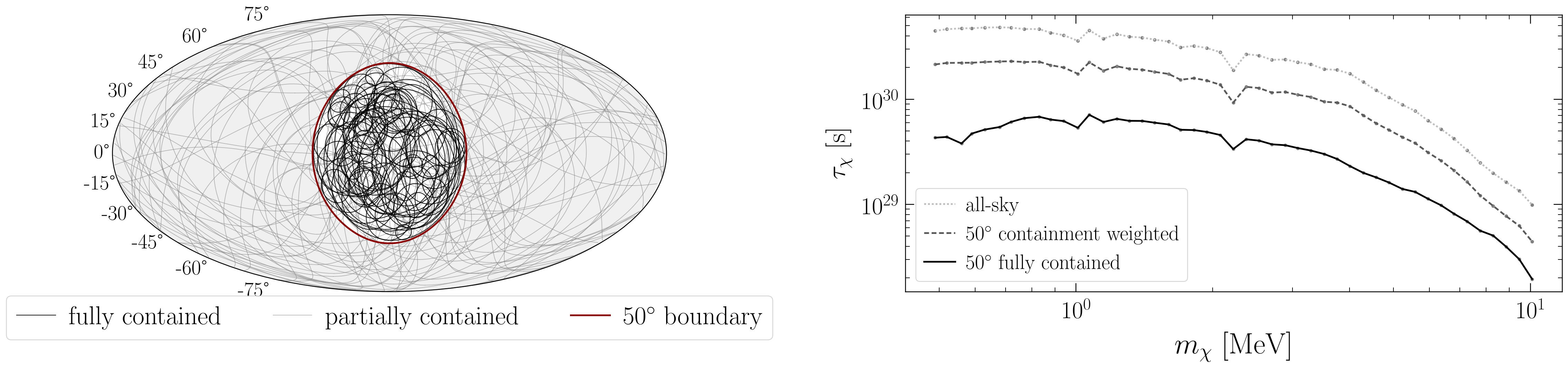}
   \caption{\emph{Left:} Compton circles projected onto the sky for a subset of simulated signal events to illustrate the containment conditions. The $50^\circ$ ROI boundary is marked by the red line, and events whose circle lies entirely inside the boundary are shown in black while those that overlap it are shown in gray. \emph{Right:} The lifetime lower limits at 95\% confidence for each containment selection, evaluated at 50 DM masses spanning the 0.4 to 10~MeV range. We use the optimal energy-only analysis here since the CDS does not contain enough counts. The all-sky selection gives the strongest limit.}
   \label{fig:roi}
\end{figure*}

Here, we implement two very simple approaches to study our $50^\circ$ ROI, roughly following the approach in Ref.~\cite{2018PhDT.......171K}. First, we apply a hard cut, considering only the DM and background events whose Compton circles are completely contained within the ROI. Second, we consider all Compton circles that intersect our ROI, weighting each event by the fraction of its Compton circle contained within the ROI. The results are shown in Fig.~\ref{fig:roi} for 50 DM masses.

This handling is very simple, as it neglects the finite width of the circle set by the angular resolution, and treats the source as uniformly distributed along each circle when a DM signal is actually concentrated toward the Galactic Center. A full treatment would use the CDS response rather than a scalar weight, which will be possible with the final version of \texttt{cosipy}. For this reason, we only present the ROI approach in this appendix as a means to illustrate how it could be pursued in principle, and we highlight that its effect is to weaken the limits compared to the all-sky limit.

\end{document}